\documentclass{ws-ijmpa}
\usepackage{url}
\usepackage[super,compress]{cite}
\usepackage{graphicx}
\usepackage[breaklinks]{hyperref}
\hypersetup{colorlinks,urlcolor=black,citecolor=black,linkcolor=black,filecolor=black}
\usepackage{breakurl}
\usepackage{colortbl} 
\usepackage[section] {placeins}
\definecolor{nodegreen}{rgb}{0,0.4,0.3}
\definecolor{weakred}{rgb}{0.8,0.6,0.6}
\definecolor{darkgreen}{rgb}{0,0.6,0} 
\definecolor{forestgreen}{rgb}{0.133,0.545,0.133}
\definecolor{purple}{rgb}{0.62745098,0.125490196,0.941176471}
\begin{document}

\markboth{J. S. SEO}
{Singularity Structure of ${\cal N}=2$ Supersymmetric Yang-Mills Theories}

%
\catchline{}{}{}{}{}
%

\title{SINGULARITY STRUCTURE OF \\ ${\cal N}=2$ SUPERSYMMETRIC YANG-MILLS THEORIES: A REVIEW\footnote{Partially based on a seminar given at Imperial College in January 2012 and a colloquium at Gwangju Institute of Science and Technology in April 2013.}}

\author{JIHYE SOFIA SEO}

\address{Ernest Rutherford Physics Department, McGill University,\\
3600 rue University, Montreal, QC H3A 2T8, Canada\\ and \\
Centre de recherches math\'{e}matiques, Universit\'{e} de Montr\'{e}al \\
C.P. 6128, succ. centre-ville, Montr\'{e}al, Qu\'{e}bec, H3C 3J7, Canada \\
jihyeseo@gmail.com}
 
\maketitle

\begin{history}
\received{Day July 2013}
\revised{Day Month 2013}
\end{history}

\begin{abstract}  
In this review, we consider the case where electrons, magnetic monopoles, and dyons become massless. Here we consider the ${\cal N} = 2$ supersymmetric Yang-Mills (SYM) theories with classical gauge groups with a rank $r$, $SU(r+1)$, $SO(2r)$, $Sp(2r)$, and $SO(2r+1)$. which are studied by 
Riemann surfaces called Seiberg-Witten curves. 
We discuss physical singularity associated with massless particles, which can be studied by geometric singularity of vanishing 1-cycles in Riemann surfaces in hyperelliptic form. 
We pay particular attention to the cases where 
 mutually non-local states become massless (Argyres-Douglas theories), which corresponds to Riemann surfaces degenerating into cusps. 
 We discuss non-trivial topology on the moduli space of the theory, which is reflected as monodromy associated to vanishing 1-cycles. 
We observe how dyon charges get changed as we move around and through singularity in moduli space.  
\end{abstract}

\ccode{PACS numbers:11.15.Tk, 11.30.Pb}

\newpage
 \tableofcontents
\section{Motivation}

Understanding of physics often advances through consideration of extreme and singular situations. 
We address lots of questions in extreme limits, with very small or very large values of density, temperature, velocity, mass etc.  These are not only for theoretical curiosity; instead often it turns out to be a golden mine for new discovery and applications. Neutron stars and big bang are at extremely high density. Superconductivity and superfluidity is studied at extremely low temperature. 
  
   Some extremes may simplify the situation enough to provide ideal setting to focus on the essence of the system. For example, ideal gas law assumes no interaction among particles. Many freshmen-level classical mechanics problems assume that friction vanishes and spring is massless.    
 
Some extremes pose us such a big challenge that it takes a paradigm shift to overcome the huddle. 
For example, consider the thought experiment in classical mechanics of the escape velocity of the satellite. If the gravity is so strong then the escape velocity approaches the speed of light. This was the first encounter (in thought) with black holes. Very careful consideration and study of the highest possible velocity, the speed of light, gave birth to special relativity, and we learned to think of space and time together as one combined object, spacetime.
    
Even in more contemporary setting, singularities in physics deserves serious attention. It may serve as a warning signal: for example, it may occur when we have integrated out massive fields which are in fact massless. UV divergences in field theory urge us to look for a better theory at higher energy. Understanding singularity is a {\it cornerstone} to solving field theory problem, just as imagining an extreme situation gives us an often correct intuition for classical mechanics problem. We often started to think about collision between particles, where their masses are equal or very different.
                 
Mathematics, especially geometry has been a faithful and fruitful language in describing physical system. Gravity - Einstein's General Relativity - is best described in the language of differential geometry. What about other forces in nature? Electromagnetic, weak, and strong forces are formulated in terms of gauge theories with gauge groups $U(1)$, $SU(2)$ and $SU(3)$ respectively. As reflected in Ref. \refcite{WuYang}, these gauge theories are well-described in another field of mathematics, so-called fiber bundle theory. 

In studying the singularity of physical system, geometry is particularly useful. Physical singularity is reflected in geometry as mathematical singularity.  There exists a famous dictionary between geometry and physics for gravitational singularity. A black hole in physics will appear as a geometric singularity, that is a puncture in a spacetime fabric. Later we will discuss physical singularity associated with both electrons and magnetic monopoles having zero mass. So far there is no known Lagrangian for this system\cite{ArgyresDouglas}. Thanks to the close relationship with geometry, however, this can be studied in terms of geometry. Recently Ref. \refcite{AMT} has revealed strange behavior of moduli space near the singularity by careful observation of geometric singularity of Seiberg-Witten curve associated to the physical system. 

In this article, we will review physical and geometric singularity of Yang-Mills theories, which have close relationship with electromagnetic and nuclear forces, but with multiple supersymmetries. The amount of supersymmetry is denoted by $\cal{N}$, the number of supercharges. Roughly it means that $2^{\cal{N}}$ particles form a set (a supermultiplet, as we will discuss in more detail later.) in which all the physical qualities are identical to one another except for the spin. We will motivate the supersymmetric Yang-Mills theories by looking at ${\cal{N}}=4$ and ${\cal{N}}=2$ theories in this introductory section, and the rest of the article will focus on ${\cal{N}}=2$ supersymmetric Yang-Mills theories.   

\subsection[${\cal{N}}=4$ SYM: gluon scattering amplitudes at hadron colliders]{${\cal{N}}=4$ SYM: gluon scattering amplitudes at hadron colliders}
One may ask
``Fine. I buy that gauge theories are important because they describe nuclear and electroweak forces. But why should anyone care about {\it supersymmetric} gauge theories, when LHC did not observe any superpartners yet?''.
First we recall that supersymmetry provides an attractive and graceful exit out of many serious paradoxical situations.
It plays a crucial role in resolving the issues of hierarchy problem, unification of coupling constants of gauge theories, mismatch of cosmological constant, etc.
In other words, supersymmetry has been a best friend to theorists, who would like to make theoretical sense and feel aesthetic harmony out of observed experimental facts.

However, supersymmetry also makes contributions for experiments. 
One of the most important inputs of supersymmetry recently is computation of scattering amplitudes of gluons. Computation is much easier when theory has supersymmetry.
At the tree level, gluon scattering amplitudes agree between supersymmetric and non-supersymmetric theories. Therefore, easier computation in supersymmetric theories can provide useful results for non-supersymmetric and more realistic theories, at least to the leading order.
Results on scattering amplitudes in maximally supersymmetric gauge theories (${\cal{N}}=4$), obtained by many string theorists, are implemented into the tools such as BlackHat\cite{BlackHat}, used by experimentalists at hadron colliders. Though massless, gluons are responsible for carrying lots of energy away from the collision process, and it is a big plus to understand their scattering amplitudes. Supersymmetric Yang-Mills theories are even more relevant in this LHC era, with or without supersymmetry detection.

 In past several years there has been a dramatic progress (almost at an exponential rate) in computation of gluon scattering amplitudes in ${\cal{N}}=4$ (maximal) supersymmetric gauge theories. 
 ${\cal{N}}=4$ supersymmetric Yang-Mills theories are special in that the 3-point function of gluons can be written down purely out of symmetry argument. Having so much supersymmetry, the theory enjoys superconformal symmetry and its dual superconformal symmetry. Conformal symmetry means one can forget about lengths. One does not even need to know the Lagrangian. One can write down S-matrix purely from the symmetry and consistency consideration, with no need for Feynman diagrams. As nicely reviewed in a recent paper Ref. \refcite{ArkaniHamed:2012nw}, to build $n$-point function the 3-point function are put together like lego blocks by amalgamation and projection operators. 
While postponing manifestation of unitarity and locality, scattering amplitudes manifest dual conformal supersymmetry and Yangian symmetry, which would remain opaque in evaluation of each Feynman diagram. Geometry is a bias-free place to look for symmetries in physics. 
 
 Gluons being massless, their 4-dimensional null (light-like) momenta enjoy amphibian lifestyle: both Lorentzian and twistor spaces\cite{Penrose:1972ia} provide a natural habitat to describe kinematics.  
 In lieu of Feynman diagrams, scattering amplitude can be organized by much simpler Hodges diagrams\cite{Hodges:2006tw,Hodges:2005bf, Hodges:2005aj} in twistor space, as nicely reviewed in Ref. \refcite{ArkaniHamed:2009si}.
 Using a higher dimensional version of Cauchy's theorem, this can be written as partial sum of residues at isolated singularities. Scattering amplitudes in maximally supersymmetric gauge theories are given as a contour integral over a Grassmannian\footnote{Grassmannian is a manifold which is a generalization of a projective space. A simple example of projective spaces is a sphere.}. Supersymmetric Yang-Mills theories thrive in a close relationship with geometry.

 Another motivation to consider ${\cal{N}}=4$ SYM is an inseparable bond between supersymmetric gauge theory (SYM) and supergravity (SUGRA). Amplitudes for SYM and SUGRA have tight kinship: supergravity amplitudes can be written in terms of SYM amplitudes, roughly speaking. One of the hot issues is the question of UV finiteness of supergravity: is supergravity a valid theory by itself, or do we must recruit string theory (or other candidates of quantum gravity) to make the supergravity consistent at arbitrarily high energy? The answer has been elusive, but supersymmetric gauge theory might be able to help. 

More excitement in ${\cal{N}}=4$ supersymmetric Yang-Mills theories can be found in Ref. \refcite{ArkaniHamed:2012nw} and its referecences. Now we will switch to less supersymmetric ones, ${\cal{N}}=2$ supersymmetric Yang-Mills theories for the rest of the review.

\subsection[${\cal{N}}=2$ SYM: massless magnetic monopole]{Motivation for ${\cal{N}}=2$ SYM: massless magnetic monopole}
Fruitful symbiosis between physics and geometry, which we observed for general relativity and ${\cal{N}}=4$ SYM, holds true for ${\cal{N}}=2$ SYM as well.
The guest of honor for ${\cal{N}}=2$ supersymmetric Yang-Mills theory is a Riemann surface. Its most famous persona is as a Seiberg-Witten curve\cite{SeibergWittenNoMatter, SeibergWittenWithMatter}: in a teamwork with 
Seiberg-Witten one-form, it encodes lots of information about the physical theories, as we will delve deeper in the rest of the review. This Riemann surface also serves as a spectral  
curve of integrable system \cite{dw}. 

Seen from the 4-dimensional field theory perspective, on which this review will mainly focus, this curve does not live inside the spacetime. One may regard it as an auxiliary object or a bookkeeping device, which happens to encode lots of useful information. This review will focus more on what to learn out of a given SW geometry, rather than how to obtain such geometry to begin with. So far the best way to understand the origin of SW geometry seems to be string theory.

In string theory settings (although we won't discuss them in depth here), the Seiberg-Witten geometry is closely related to extended objects in string theories and M-theory. Ref. \refcite{wittenM} interprets that the 4-dimensional field theory comes from wrapping M5-brane on the Seiberg-Witten curve. In M-theory, M5 brane is a solitonic object spanning 5 spatial and 1 temporal directions, carrying a conserved charge. 
Just as we consider world-line of a point particle traveling in time, we can consider 6-dimensional theory on the world-volume spanned by time evolution of M5-brane.
However if we let M5-brane to wrap a 2-dimensional Riemann surface and further assume that the Riemann surface is small compared to other directions in the spacetime, then we will have only 4 remaining directions effectively. This type of 4-dimensional theories are discussed in Refs. \refcite{wittenM,gaiotto}.  
There are also interpretations of Seiberg-Witten curve in terms of non-critical and anti-self dual strings in type IIA, IIB, and heterotic string theories, again appearing as wrapping extended solitonic objects on appropriate cycles, as reviewed in Ref. \refcite{LercheReview}. 

Thanks to Seiberg-Witten theory\cite{SeibergWittenNoMatter, SeibergWittenWithMatter}, many ${\cal{N}}=2$ supersymmetric Yang-Mills theories can be equivalently written as a Riemann surface (written down as a format of hyperelliptic curve) called Seiberg-Witten curve with a one-form. 
Studying singularity on geometry-side provides us a powerful microscope probing singularity of physical theory. Degeneration and monodromy of hyperelliptic curves translates into massless fields and their dyon charges in Seiberg-Witten theories\footnote{A dyon refers to a particle which potentially carries both electric and magnetic charges.}. 
A higher singularity coming from collision of milder singularity gives us an exotic theory with massless electron and massless monopole, so-called Argyres-Douglas theory. It defies Lagrangian description: when Lagrangian mechanics turns its back on us, we have all the more reason to seek the friendship with geometry.

Let us pause for a moment and remind ourselves why we need to study magnetic monopoles, especially why the light ones. Despite many experimental claims and findings, magnetic monopole (massive or massless) is something we have not observed in a concrete manner yet. 
In the set of Maxwell's equations, magnetic monopoles naturally arise if one tries to manifest the hypothetical electro-magnetic symmetry, and introduces magnetic sources just like electric sources. There are two main reasons why the magnetic monopole must exist beyond a theorist's fanciful imagination, as pedagogically reviewed in Ref. \refcite{PhysRevD.86.010001} by D. Milstead and E. J. Weinberg. Grand Unified Theory (GUT) (of electromagnetism and nuclear forces) predicts existence of magnetic monopole as shown in Refs. \refcite{'tHooft,Polyakov}.  The mere existence of magnetic monopole explains and necessitates quantization of electric charge.\cite{Dirac} 

For the lack of experimental evidence of monopole, we tend to blame its high mass. In the present Universe, we expect the magnetic monopole to be very heavy - its mass energy is near that of a bacterium or near kinetic energy of a running hippo, which is a lot larger compared to other elementary particles. Big bang also provides an excuse for its absence in experimental data, arguing that the finite number density of magnetic monopole diluted out as the universe evolves. However if we trace back the history of universe, then spontaneously broken symmetries (such as GUT and supersymmetry) are restored and magnetic monopole might have been not so heavy. Particles which are partners under supersymmetry and electromagnetic duality can be thought of babies who were born as identical twins, but as time goes on, who grow into adults with different physical qualities.
              
  Supersymmetric Yang-Mills theory is a promising place to learn about quantum field theories. First, supersymmetry allows computation. Second, some properties we find in supersymmetric theories often still hold even in non-supersymmetric quantum field theories. In some sense supersymmetric field theories provide fruitful and fertile toy models to learn about realistic quantum field theories.

\subsection{Plan of the review \label{Plan}}

In section \ref{essenSec}, we revisit essential elements of ${\cal{N}}=2$ supersymmetric Yang-Mills theories and Seiberg-Witten geometry, and prepare ourselves with necessary tools for studying singularity. In section \ref{firstlook}, we study Seiberg-Witten curves for $SU(r+1)$ and $Sp(2r)$ SYM and discuss the  root structure of those families of hyperelliptic curve, in preparation for section \ref{monodromySec}
where we compute dyon charges of massless states. Section \ref{ArDo} deals with higher singularity with massless electron and monopole (Argyres-Douglas theories) among ${\cal N} = 2$ supersymmetric Yang-Mills theories with classical gauge groups $SU(r+1)$, $SO(2r)$, $Sp(2r)$, and $SO(2r+1)$.
  In section \ref{doublediscsection}, we revisit the tools to capture singularity to learn more about the singularity structure.   
We conclude with open questions in section \ref{conclusion}.

  \section{Essentials of ${\cal{N}}=2$ SYM and Seiberg-Witten Geometry \label{essenSec}}
 
Many wonderful reviews\cite{BilalReview, LercheReview, AlvarezGaumeHassanReview, Argyres:1998vt} exist on ${\cal{N}}=2$ SYM, while this review is more focussed on the Seiberg-Witten geometry and singularity of those theories. Here we will only pinpoint the aspects that are necessary for understanding the main idea of the review. The Lagrangian of 
  ${\cal{N}}=2$ supersymmetric Yang-Mills theory can be written elegantly in ${\cal{N}}=1$ or ${\cal{N}}=2$ superspace language, where supersymmetry is more manifest. Here we will write it down in a spelled-out fashion in 4-dimensional spacetime language as below: 
\begin{eqnarray} {\cal{L}} &=&{\color{red}    - \frac{1}{g^2} \int d^4 x {\rm Tr}   \left[ \frac{1}{4} F_{\mu \nu} F^{\mu \nu} \right]  }+ {\color{purple} \frac{\theta}{32 \pi^2} \int d^4 x {\rm Tr} \frac{i}{4} F_{\mu \nu} \tilde{F}^{\mu \nu}} \nonumber \\  
& & {\color{forestgreen}  -  \int d^4 x  \frac{1}{ 2}  {\rm Tr} \left[ \phi^+ , \phi \right]^2 }+{\color{blue} {\mathrm{(fermions)}}}. \label{N2Lag}
\end{eqnarray}    
Here $A_\mu$ and $\phi$ transform as a vector and a scalar (respectively) of Lorentz group $SO(3,1)$ of spacetime. Both are adjoint representation of the gauge group $G$. The choice of gauge group also determines structure constants $f_{abc}$ and generators $T_a$,  
\begin{eqnarray} A_\mu &=& A^a_\mu T_a, \quad [T_a, T_b]=f_{abc} T_c , \nonumber \\
  F_{\mu\nu} &=&\partial_\mu A_\nu -\partial_\nu A_\mu - ig [A_\mu,A_\nu], \quad   \tilde{F}^{\mu\nu} =\frac{1}{2} \epsilon^{\mu\nu\rho\sigma} F_{\rho \sigma}  \end{eqnarray} and $g, \theta $ are real-valued coupling constants.  
   If we did not have supersymmetry, then only the {\color{red} first term} of Eq. \eqref{N2Lag} would appear in the Lagrangian, as the {Yang-Mills} action. For example, the action for weak and strong nuclear forces can be written down by choosing $G=SU(2), SU(3)$ and taking only the first term of Eq. \eqref{N2Lag}. Having ${\cal{N}}=2$ extended supersymmetry dictates that scalar, spinors, and vector must transform together forming a ${\cal{N}}=2$ supermultiplet\footnote{More specifically it is called ${\cal{N}}=2$ vector multiplet for it contains a vector. The space whose coordinates are the scalar components of vector multiplets is called a Coulomb-branch of moduli space. The scalars of another ${\cal{N}}=2$ supersymmetry representation, hypermultiplet, form a Higgs moduli space. A supermultiplet is a representation of a supersymmetry algebra. More details are given in excellent reviews, Refs. \refcite{AlvarezGaumeHassanReview,Argyres:1998vt}.}. Instead of keeping track of all component fields in a given supermultiplet, we can save our effort and restrict our attention to the term one of them only. Here it is enough to consider the part with the scalar $\phi$ only which is the {\color{forestgreen} third term} of Eq. \eqref{N2Lag}, focussing on vacuum structure of a ${\cal{N}}=2$ supersymmetric Yang-Mills theory.
   In Eq. \eqref{N2Lag}
 the {\color{purple} second term} gives {instanton number} and the {\color{blue} last term} denote terms involving {fermions}.    
      
However we won't make usage of this Lagrangian any more in this review, because Seiberg and Witten proposed another `geometric' way to study 
 ${\cal N} = 2$ SYM theories\cite{SeibergWittenNoMatter, SeibergWittenWithMatter}, which we now turn to.
 
\subsection{Review of ${\cal N}=2$ Seiberg-Witten theory and geometry \label{ReviewN2}}

In late 1990s, Seiberg and Witten made a profound discovery on ${\cal N} = 2$, $d = 4$ supersymmetric Yang-Mills theory with gauge group $G=SU(2)$\cite{SeibergWittenNoMatter, SeibergWittenWithMatter}, 
giving a huge impact both on physics and mathematics. After two dozen years, Gaiotto blew a new life into research on ${\cal N} = 2$ superconformal theories recently, by discovering a plethora of new theories with often surprising features, which can be all lego-ed from simple building blocks\cite{gaiotto}\footnote{All these methods study theories with low energy effective action. This review also deals with those only.}. 
   
\begin{figure}[htb]
   \begin{center}
        \includegraphics[width=\textwidth]{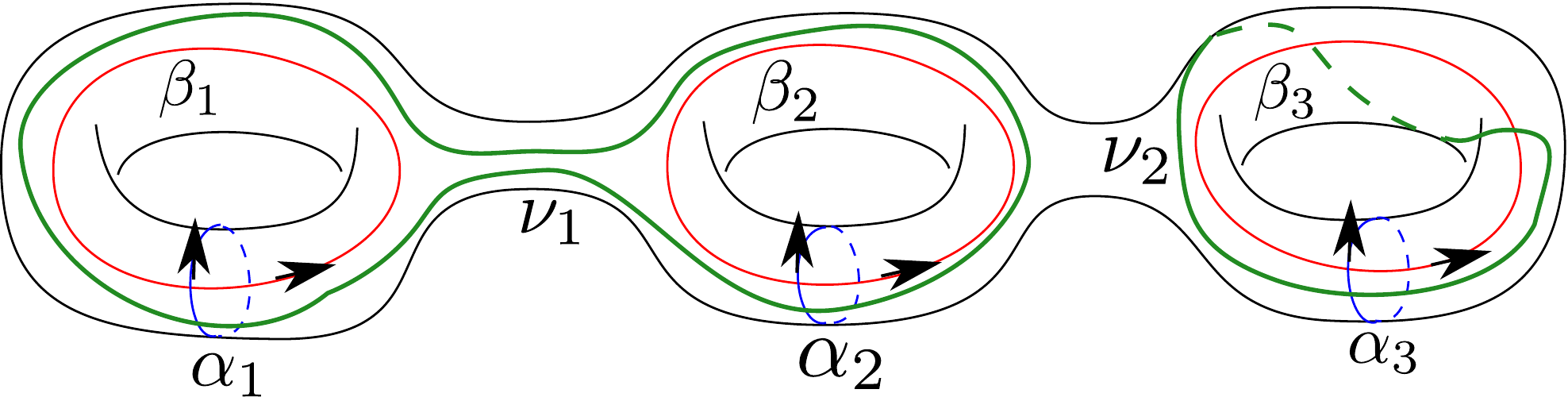}
   \caption{Various 1-cycles and their symplectic basis for a Riemann surface of genus 3}
      \label{nu}
    \end{center}
   \end{figure}

Simply speaking, Seiberg and Witten proposed a powerful dictionary between physics and geometry for ${\cal{N}}=2$ theories.      
 Seiberg-Witten geometry comes in package with Seiberg-Witten (SW) curve and Seiberg-Witten (SW) differential 1-form. 
 The SW curve is a complex curve, or a real 2 dimensional Riemann surface, whose genus is equal to the rank $r$ of the gauge group (such as $SU(r+1)$, $Sp(2r)$, $SO(2r)$, and $SO(2r+1)$) for the supersymmetric Yang-Mills theories (that is, with no matters added). It is also equal to the complex dimension of moduli space. Recall that the moduli are to be understood as parameters controlling the theory and the subsequent SW geometry. If the gauge group had rank 3, then the corresponding SW curve may look like the Riemann surface in Fig. \ref{nu}, with genus 3.

        \begin{equation} 
\begin{tabular}{|ccc|}
\hline Physics & $\leftrightarrow$ &  Geometry  \\ \hline
\hline  supersymmetric Yang-Mills theory & $\leftrightarrow$ & Riemann surface  \\  \hline
rank of gauge group  & $\leftrightarrow$ & genus  \\ \hline
(BPS) particles  & $\leftrightarrow$ &  (some) 1-cycles \\ \hline
   \end{tabular}     \label{tabledic}
\end{equation}

\begin{figure}[htb]
   \begin{center}
        \includegraphics[width=\textwidth]{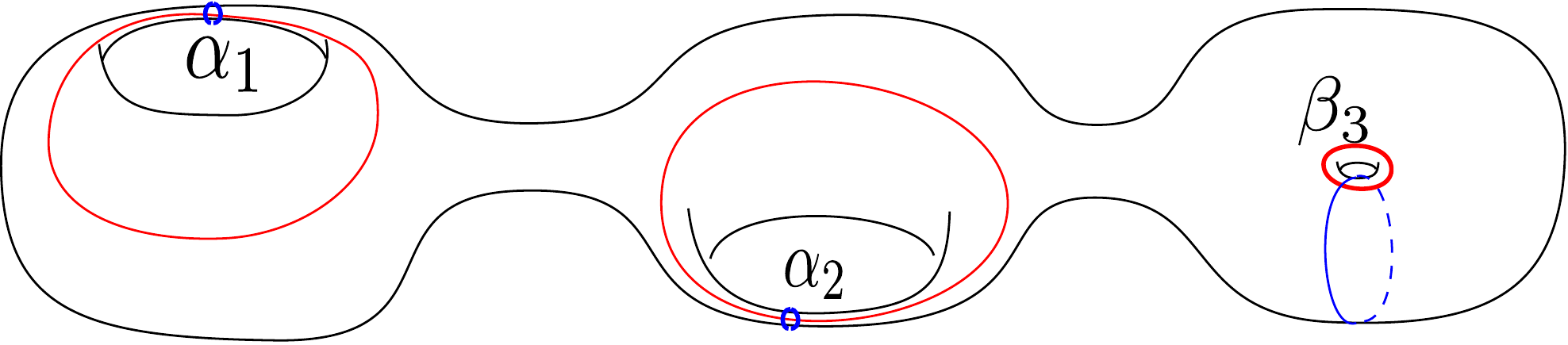}
    \end{center}
    \caption{Vanishing 1-cycles of genus-3 Riemann surface. All these 3 cycles are mutually local, since intersection numbers all vanish.}
    \label{local}
   \end{figure}

\begin{figure}[htb]
   \begin{center}
   \includegraphics[width=\textwidth]{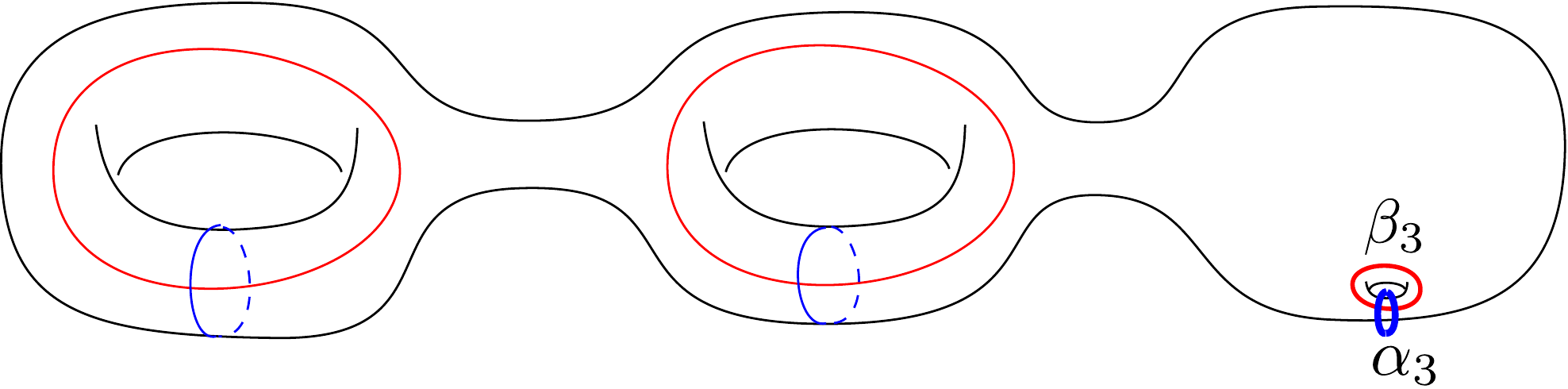}
       \caption{Mutually non-local vanishing cycles of genus-3 Riemann surface. Their intersection number is non-zero.}
    \label{nonlocal}
     \end{center}
   \end{figure}

 In pure Seiberg-Witten theory the dimension of the moduli space (or the number of moduli/parameters) is also equal to the genus\footnote{One may introduce matters into the Seiberg-Witten theory: By {\it pure} SW theory, they mean lack of matter, and it will be our focus on this review.}, which, in turn, is equal to the rank of the gauge group. At a generic point in the moduli space, the SW curve is smooth and all the 1-cycles are non-vanishing as in {Fig.~\ref{nu}}. However, we could move to a less generic location in the moduli space where we have vanishing 1-cycles as in Fig. \ref{local} and Fig. \ref{nonlocal}. 

Note that on a SW curve, we can draw various 1-cycles as in {Fig.~\ref{nu}}. 
Here we have chosen a particular set of 
symplectic basis 1-cycles, ${\color{blue}\alpha_i}$'s and ${\color{red}\beta_i}$'s. The only rule to keep for the choice of symplectic basis cycles is that the intersection numbers must satisfy:
 \begin{equation}
 {\color{blue}\beta_i } \circ {\color{red}\alpha_j} =\delta_{ij}. \label{intersectsymp}
\end{equation}
The intersection number is an anti-symmetric ($   {\color{red}\alpha_j} \circ {\color{blue}\beta_i } =-\delta_{ij}$) and bilinear (linear dependence on both arguments) operation among 1-cycles\footnote{For later convenience, the choice of overall sign for intersection number chosen to match that of Ref. \refcite{LercheReview} and is opposite of that of Ref. \refcite{SD}.}. Each 1-cycle has an orientation (as seen by the arrow in figures), and the intersection number comes with a sign\footnote{This is an algebraic intersection number, as opposed to a geometric intersection number.}.

As anticipated from Eqn. \eqref{tabledic}, 1-cycles of the Riemann surface correspond to physical particles\footnote{Only some of 1-cycles, which pass the test of wall-crossing formulas, correspond to stable BPS (supersymmetric) particles. Our main focus here is for massless states. With an assumption that the massless states are stable, we can consider 1-1 mapping between 1-cycles and BPS particles, with restriction to the massless sector.}, and a choice of symplectic basis 1-cycles assigns electric and magnetic charges to the particles.
For given $i$, ${\color{blue}\alpha_i}$ and ${\color{red}\beta_i}$ denote electric and magnetic charge (respectively) for the $i$'th $U(1)$ inside the gauge group $G$. 

However, the choice is certainly not unique, and we could modify the choice by 
\begin{equation} 
{\color{blue}\alpha_i^{\prime}} \equiv {\color{red}\beta_i}, \qquad {\color{red}\beta_i^{\prime}}  \equiv - {\color{blue}\alpha_i} \label{emduality} 
\end{equation}
 for given $i$ only, and this still preserves the symplectic property of Eqn. \eqref{intersectsymp}. In physics this corresponds to the electromagnetic duality on $i$'th $U(1)$ charge. 
 Another important fact is that the intersection number (being a scalar) is invariant under electromagnetic dualities of Eq. \eqref{emduality}, and in general, under symplectic transformation (re-choice of symplectic basis 1-cycles). 
 
 Some of 1-cycles correspond to physical states (stable BPS/supersymmetric dyon), with quantized {\color{blue}electric} and {\color{red}magnetic} charges. As shown Fig. \ref{nu}, any 1-cycle can be written in terms of basis 1-cycles ${\color{blue}\alpha_{i}}$'s and ${\color{red}\beta_i}$'s with integer coefficients, with dyonic charges superposed. These integer coefficients exactly correspond to amount of electric and magnetic charges of each $U(1)$. Two 1-cycles $\nu_{1}$ and $\nu_{2}$ in Fig. \ref{nu} can be written as follows:
    \begin{equation}
  {\color{forestgreen}\nu_1} = {\color{red}\beta_{1}} +{\color{red}\beta_{2}} , \qquad
    {\color{forestgreen}\nu_2} = -{\color{blue}\alpha_{3}} +{\color{red}\beta_{3}}. \label{dyonchargecycle}
    \end{equation} 
 Physical interpretation of this would be that, if they corresponded to BPS dyons, then the first object (${\color{forestgreen}\nu_1}$) behaves as a magnetic monopole for both the first and the second $U(1)$s, and the second object ($ {\color{forestgreen}\nu_2}$) carries the same charge as a bound state of a positron and magnetic monopole of the third $U(1)$.
 
Seiberg-Witten geometry contains lots of (if not all) information about the theory. The physical information is stored not only in SW curves, but also in the SW 1-form. The SW curve and SW 1-form work together, and without each other they lose meaning, just like a needle and a thread. The SW curve provides 1-cycles over which to integrate the SW 1-form. Then we obtain complex number which is meaningful physically (central charge). By integrating Seiberg-Witten differential 1-form $\lambda_{\rm SW}$ over 1-cycle $\nu$, we obtain a complex number. For the purpose of this review, we are only interested in its magnitude, which is the mass of the particle
\begin{equation} M_{\nu}= \left| \oint_\nu \lambda_{\rm SW} \right|. \label{Mlambda} \end{equation}
 Since we are focussing on physical singularity associated with massless particles, Eqn. \eqref{Mlambda} provides the most important piece of information for the purpose of this review, among what we learn from the Seiberg-Witten geometry. 
Assuming $\lambda_{\rm SW}$ is free of delta-function behavior, 
vanishing of 1-cycle $\nu$ signals existence of massless BPS state (with dyonic charge given by $\nu$) since its mass given in Eqn. \eqref{Mlambda} vanishes\footnote{If the integrand $\lambda_{\rm SW}$ has a delta-function type singularity, integrating it over an infinitesimal interval may give a finite value to Eqn. \eqref{Mlambda}.}. Therefore, study of vanishing 1-cycles can teach us about massless BPS states in the system. 
We therefore assume that:

\vskip.1in

\framebox[1.1\width]{Singularity loci of SW curve $\subset$  Singularity loci of SW theory.}    
 
 \vskip.1in
  
  Now let us imagine tuning various parameters (moduli) for the gauge theory of Fig. \ref{nu}, to force some 1-cycles to vanish. In Fig. \ref{local}, we have three vanishing 1-cycles $  {\color{blue}\alpha_1}, {\color{blue}\alpha_2},$ and ${\color{red}\beta_3}$ which do not intersect each other. They correspond to an electron with respect to the first $U(1)$, another electron with respect to the second $U(1)$, and a magnetic monopole with respect to the third $U(1)$. All three particles are massless. 
 If we operate an electromagnetic duality on the third $U(1)$, then the third particle will be renamed into a massless electron with respect to the third $U(1)$. All the massless particles are mutually {\it local}, in that they can be treated as pure electrons (carrying no magnetic charge) in some choice of symplectic basis 1-cycles (i.e. after a certain series of performing electromagnetic dualities). This is possible only because (if and only if, in fact) the corresponding 1-cycles have vanishing intersection number with one another. An equivalent mathematical statement is this: if all the 1-cycles in a certain set have zero intersection number with one another, then they can be written in terms of linear combination of $\alpha_i$'s with no need for $\beta_i$ terms. We will soon explain why we call them {\it local}, (near Eqn. \eqref{PLformula}) after explaining non-locality now.
 
      Two 1-cycles ${\color{blue}\alpha_3}$ and ${\color{red}\beta_{3}}$ vanish in {Fig.~\ref{nonlocal}}, and they correspond to an electron and a magnetic monopole, both charged with respect to the third $U(1)$, and with zero mass. No matter how one may try to redefine electric and magnetic charges by electromagnetic dualities and so on, it is never possible to make both of them into electric particles at the same time. That is because the two vanishing 1-cycles have non-zero intersection number $   {\color{blue}\alpha_3} \circ {\color{red}\beta_{3}}=-1\ne 0$, regardless of choice of symplectic bases.
If a set of 1-cycles were able to be written as electric particles (in terms of $\alpha_i$'s only), then they must have had zero intersection number with one another.  
 
We will pause briefly here to explain naming of locality versus non-locality for vanishing 1-cycles (massless particles, equivalently). As nicely reviewed in Ref. \refcite{LercheReview}, 1-cycles transform under monodromy action, as one moves around on a non-contractible loop, surrounding a singularity, in moduli space (changing the moduli values accordingly). If the singularity is where a 1-cycle $\nu$ vanishes, then the other 1-cycle $\gamma$ gets transformed according to this Picard-Lefshetz formula, as explained in Ref. \refcite{LercheReview}
\begin{equation}
M_\nu : \gamma \rightarrow \gamma - (\gamma \circ \nu) \nu . \label{PLformula}
\end{equation}
For a 1-cycle $\gamma$ which does not intersect with $\nu$ i.e. $\gamma \circ \nu=0$, no change will be made on $\gamma$ under monodromy. (However, any 1-cycles which intersect with $\nu$ will be shifted as one goes around the singularity loci associated with vanishing of the 1-cycle $\nu$, as we will explain below.)

This gives a motivation for the naming: if two cycles have zero intersection number, when one vanishes, the other cycle does not get affected. Or in physics language, two particles can be written as purely electric ones at the same time. When one becomes massless the other does not change its charges. In some sense, they do not need to care about each other, and they are mutually {\it local}. It is possible to write down Lagrangian for those theories, by adding each local pieces.

However, now assume that two cycles have non-zero intersection number. When one vanishes, the other cycle receives a monodromic shift. 
In physics language, the two particles are mutually {\it non-local} and they cannot be written as purely electric ones at the same time. 
When one particle becomes massless, it is ambiguous how to assign charge to the other particle. 
The dyon charge of the second particle is not a single-valued function of moduli near the singularity locus where the first particle becomes massless. 

In general, there is no known Lagrangian for these systems.  
      However, in Ref. \refcite{ArgyresDouglas}, this exotic theory (so-called Argyres-Douglas theory) has been discovered and studied, inside moduli space of $SU(r+1)$ SW theories. Since there is no Lagrangian description yet (if not never), studies are conducted by careful analysis of scaling dimensions near the singularity loci of Seiberg-Witten geometry which can be written as a hyperelliptic curve equipped with 1-form. Recent key developments in this direction can be found in Refs. \refcite{AMT,GST} among others. Now we turn to review geometry of hyperelliptic curves.
      
     \subsection{Review of hyperelliptic curves \label{GeomReview} $y^2=f(x)$}
     So far in this section, we discussed Riemann surface with 1-cycles which potentially could collapse. For the purpose of this review, the Riemann surfaces of our interest can be written as an algebraic variety given by $y^2=f(x)$, which include hyperelliptic curves. 
     
     Since we need to deal with singularity as well, let us begin by recalling a few fundamental facts about singularity of algebraic varieties. Let us consider an algebraic variety given by $F(x,y,z,\ldots)=0$. This is an object embedded inside a bigger space, {\it ambient space} whose coordinates are $x,y,z,\ldots$. It is singular if exterior derivative $dF=0$ vanishes, or in other words if all the partial derivatives vanish, namely $\frac{\partial F}{\partial x}    =  \frac{\partial F}{\partial y}= \cdots=0$.  
     
       The exterior derivative $d$ is written in terms of the partial derivatives with respect to all the coordinates of the ambient space. Since the Riemann surface is embedded in an ambient space whose coordinates are $x$ and $y$, the exterior derivative is given as  \begin{equation}d= dx \frac{\partial  }{\partial x} + dy \frac{\partial  }{\partial y}. \end{equation}  Later, we will consider algebraic variety embedded inside the moduli space whose coordinates are complex-valued moduli $u_i$'s, then the exterior derivative will accordingly be $d= \sum_i du_i \frac{\partial  }{\partial u_i} $.
    
     In simpler words, at each point on a given surface (algebraic variety) embedded in a bigger space, {\it an ambient space}, we can consider tangent space. However, if the surface develops singularity, then the tangent space suddenly changes there ($dF=0$). Singularity of an algebraic variety (super-elliptic curve) 
     \begin{equation}
     F \equiv y^n-f(x)=0, \quad n \ge 2  \label{suel}
     \end{equation}
     is given by having $\frac{\partial F}{\partial x} = -\frac{\partial f}{\partial x}=0 $ and $ \frac{\partial F}{\partial y}=n y^{n-1}=0$. Therefore the singularity is at where $y=0=f(x)=\frac{\partial f}{\partial x}$. In order for $f(x)$ and $\frac{\partial f}{\partial x}$
 to have a common root, it is equivalent to demanding $f(x)$ to have a degenerate root. We will now see that it happens if and only if $f(x)$ has vanishing discriminant $\Delta_x f =0$. 
 
      {Discriminant} of a polynomial $f_n({\color{red}x}) =  \prod_{i=1}^n ({\color{red}x}-e_i)  $ is given in terms of its roots as 
\begin{equation}
\Delta_{\color{red}x} \left( f_n(x) \right) =   \prod_{i<j} (e_i-e_j)^2.  \label{DiscDef}
\end{equation} Vanishing of Eqn. \eqref{DiscDef} is equivalent to existence of repeated roots.
The number of identical roots is called the degeneracy, multiplicity of zero, or
order of vanishing.  

 A subscript for the discriminant symbol denotes which variable we take discriminant with respect to. This will be useful when we have a polynomial in multiple variables. For example, we will first discuss discriminant with respect to $x$, in the ambient space whose coordinates are $x, y$, in which Riemann surface is embedded. Next we will discuss algebraic variety defined inside the space of moduli, the parameters which control properties of Riemann surface. Then we will take discriminant with respect to one of the variables in the moduli space.
 
Smooth hyperelliptic curve is defined (similarly to Eqn. \eqref{suel}) as 
a complex curve embedded in an ambient space whose coordinates are two complex variables $x, y \in \mathbb{C}$ satisfying the equation 
\begin{equation} y^2 = f_n(x)= \prod_{i=1}^n ({ x}-e_i), \quad n > 4  \end{equation}
where complex parameters $e_i \in \mathbb{C}$'s are all distinct from each other ($ e_i \ne e_j $ for $i\ne j$).
This gives a double-sheet fibration of $x$-plane with multiple ($n$ for even $n$, $n+1$ for odd $n$) separate branch points for a Riemann surface. For generic value of $x$, $y = \pm \sqrt{f(x)}$ has two choices for sign, therefore creating double-sheet. We can choose the upper and lower sheets to satisfy $y = \sqrt{f(x)}$ and $y = - \sqrt{f(x)}$ respectively. If $x=e_i$, then $y=0$ and branch points will be formed.
Obviously, $n$ branch points are at each $e_i$'s. If $n$ is odd, then we have an extra branch point at $x=\infty$ because of an extra monodromy of
$y^2=f_{{\rm odd}\ n}(x)$ there: As $x$ rotates by $2\pi$, $y$ changes its sign if and only if $n$ is odd.
In the double-sheet fibration picture, each upper and lower $x$-plane can be thought as a sphere (with compactification at infinity). Each pair of branch points can be considered as a tube (cylinder) connecting two spheres. Therefore, the genus is $g=\left[ \frac{n-1}{2} \right]$\footnote{The square bracket $[ \ ]$ denotes the floor function. $[x]$ is the largest integer which satisfies $[x] \le x$.}. When $n=3,4$ as in rank 1 SW curve, then this formula asserts that $g=1$. 
Recalling the definition of discriminant, demanding $ e_i \ne e_j $ for $i\ne j$ guarantees smoothness.

However, we want to consider possibility of singularity and vanishing 1-cycles (for example coming from shrinking of the cylinder connecting two spheres). Therefore for the review, we will extend the definition of hyperelliptic curve as a complex variety given by an equation 
\begin{equation}
 y^2 = f_n(x)= \prod_{i=1}^n ({ x}-e_i), \quad x, y, e_i \in \mathbb{C},  \quad n > 4. \label{hecd}
\end{equation}
 The only difference is that we no longer demand the $e_i$'s to be distinct.
Of course physicists are already motivated to look at singular curve due to massless states, as explained near Eqn. \eqref{Mlambda}. 
As we explained near  
 Eqn. \eqref{PLformula}, even if one is only interested in smooth hyperelliptic curves, the moduli space has nontrivial topology where 1-cycles transform under monodromy as we go around a non-contractible loop in a moduli space.
 
Therefore we consider potentially singular, hyperelliptic curve, as defined in Eqn. \eqref{hecd}, which is a double-sheet fibration of $x$-plane with $2g+2$ branch points.
When roots degenerate, the curve degenerates.
 
   Because of the squaring in Eqn. \eqref{DiscDef}, the discriminant is symmetric among the roots $e_i$'s. Therefore, the discriminant can also be expressed in terms of the coefficients of the polynomial $f_n(x)$ (therefore moduli). 
     
For an algebraic curve given as $y^2=f(x)$, by a discriminant of the curve, we mean discriminant $\Delta_x f$. Singularity of the curve is captured by colliding roots on the $x$-plane, at vanishing discriminant $\Delta_x f$. Note from right-hand side of Eqn. \eqref{DiscDef} that 
           $\Delta_x f$ has no dependence on $x$ or $y$. Demanding $\Delta_x f=0$ only gives one complex condition inside moduli space. Therefore vanishing discriminant loci is an algebraic variety inside moduli space with complex codimension one. By {\it codimension}, we mean an embedded object has smaller dimension than the ambient space by it, so it is equal to the number of independent conditions imposed.
            
  Existence of degenerate root of $f(x)$ signals singularity: there exists a vanishing 1-cycle. Various 1-cycles of $y^2=f(x)$ can vanish as we let the roots of $f(x)$ to degenerate, as depicted in Fig. \ref{local}. A `donut' can degenerate into a thin `ring' with shrinking $\alpha$ cycle, or into a fat `bagel' with shrinking $\beta$ cycle. 
 When the degeneracy of the root is high (3 or larger), then it signals that there are multiple vanishing 1-cycles which have non-zero intersection numbers with one another, as depicted in Fig. \ref{nonlocal}. Both $\alpha$ and $\beta$ cycles shrink together for the same `donut', forming a cusp (for example a `croissant').
This corresponds to an Argyres-Douglas theory, with {\emph {mutually non-local}} massless dyons. 

Each time we demand a branch point to collide with another, we are using up a degree of freedom. Therefore, by demanding 3 branch points to collide all together, we use up two degree of freedom. This means that the Argyres-Douglas theory occurs in complex codimension 2 loci in moduli space. When the degeneracy of branch point is maximized, we call it the maximal Argyres-Douglas point. 
  For pure SYM case with gauge group of rank $r$, we have $r$ degrees of freedom, and we can bring $r+1$ branch points together to form maximal Argyres-Douglas {\it isolated} points in the moduli space\footnote{By isolated points, we roughly mean that they form a discrete set and are separated. For example, the points do not congregate to form a line or a plane. \label{isolated}}.
 
      Argyres-Douglas theory contains mutually non-local massless particles, and they occur in a complex codimension two loci of pure Seiberg-Witten theories. Due to presence of massless electron and magnetic monopole, Lagrangian cannot be written down. Here hyperelliptic form of Seiberg-Witten curve forms a cusp-like singularity (or worse), at complex codimension 2 loci of moduli space of ${\cal N} = 2$ theories.

\section{First Look at Seiberg-Witten Curves for $SU(r+1)$ and $Sp(2r)$ \label{firstlook}}
Among hyperelliptic curves given in Eqn. \eqref{hecd}, 
here we will consider a few families only, which are Seiberg-Witten curves for SYM with $SU(r+1)$ and $Sp(2r)$ gauge groups. In the parameter space of the hyperelliptic curves, these will form subspaces with dimension almost halved\footnote{Recall from the definition of hyperelliptic curve in Eqn. \eqref{hecd} that we allow the coefficient of each power of $x$ to vary, all independently from each other. For most generic hyperelliptic curve, the overall power of $f(x)$ equals to the number of parameters. By assigning the hyperelliptic curve a role of SW curve for certain gauge groups, we no longer have the full freedom of varying all of them. As we will see soon the number of independent parameters is $r$ while $f(x)$ has the power $2r+1, 2r+2$ etc.}. Here we will focus on `root structure', in other words, potential degeneracy of branch points. For both cases, we note that the curve is factorized into two polynomials which never share roots. Branch points will be divided into two mutually-exclusive sets where multiplicity may happen only within each group. Each set of branch points will be assigned with a name and a color, therefore enabling bi-coloring (green and purple) of coming figures in this review.

\subsection{Seiberg-Witten geometry for pure ${\mathcal N}=2$ $SU(r+1)$ theories \label{suSec}} 

 The 
SW curve and SW 1-form for pure $SU(r+1)$ of Ref. \refcite{KLYTsimpleADE} are rewritten as
  \begin{equation}  
 y^{2}=f_{SU(r+1)} = f_{+} f_{ -} , \quad   \lambda_{\mathrm SW} = -dx \log \left( - \frac{1}{2}(f_+ + f_-) - \sqrt{f_+ f_-}\right), \label{surcurve1form}
 \end{equation}
where $f_{\pm}$ are given in terms of $r$ gauge invariant complex-valued moduli $u_i$'s as:
  \begin{equation}  
  f_{\pm} \equiv  x^{r+1} + \sum_{i=1}^r u_i x^{r-i}   \pm \Lambda
^{ r+1}.  \label{fpm}
\end{equation}   
Note that we are not allowing the full $2r+1$ degrees of freedom of hyperelliptic curves of Eqn. \eqref{hecd}. Instead we get to vary $r$ (same as the rank of the gauge group) moduli $u_i$'s only while $\Lambda$ is a fixed non-zero constant.

In discussing singularity of hyperelliptic curve, we will consider collision among the branch points of $f(x)$. Therefore, it is convenient to give names to the roots of $f_{\pm}$, as 
 \begin{equation}f_{+} \equiv \prod_{i=0}^{r} (x-P_i), \qquad f_{-} \equiv \prod_{i=0}^{r} (x-N_i). \label{fpmPN}\end{equation}
 
Since $\Lambda\ne 0$, $f_{\pm}$ can never vanish at the same time. Therefore $f_+$ and $f_-$ can never share a root, and there is no vanishing 1-cycle mixing these two sets of roots. 
More explicitly, the discriminant of the $SU(r+1)$ SW curve factorizes into\cite{SD}
\begin{equation}\label{sufactordisc} \Delta_x f_{SU(r+1)}=(2\Lambda^{ r+1})^{2r+2} \Delta_x f_{+} \Delta_x f_{-}.
\end{equation}
In other words, in order for $f_{SU(r+1)}(x)$ to have a degenerate root, $f_+$ or $f_-$ itself should have a degenerate root. 
This justifies binary color coding in figures for branch points and vanishing 1-cycles. 
On the $x$-plane, only $P_i$'s (or $N_i$'s) can collide among themselves.

 At discriminant loci $ \Delta_x f_{SU(r+1)}=0$ and near the corresponding vanishing 1-cycle, 
the 1-form of \eqref{surcurve1form} is regular
\begin{equation}
\lambda_{\mathrm SW} = -dx \log \left(       \pm   \Lambda^{r+1}   \right), \qquad {\mathrm{near}}  \ f_\pm = \Delta_x f_{\pm}=0,
\end{equation} confirming that the singularity of the SW curve is indeed the singularity of the SW theory, as promised earlier above Fig. \ref{local}.

\subsection[Seiberg-Witten curve for pure $Sp(2r)$ theories and root structure]{SW curve for pure $Sp(2r)$ theories and root structure \label{sprootstructure}}

Now consider a slightly different hyperelliptic curve,
\begin{equation}    
 y^{2}=f_{Sp(2r)}= f_C f_Q , \quad  \lambda   =  a \frac{dx}{2\sqrt{x}}  \log \left( \frac{   x f_C + f_Q+ 2\sqrt{x} y   }{  x f_C + f_Q -2\sqrt{x} y  } \right), \label{sprcurve1form}
\end{equation}    
with $f_C$ and $f_Q$ defined as:
\begin{equation}   
f_C  \equiv x^{r } + \sum_{i=1}^r u_i x^{r-i}, \qquad f_Q \equiv  x f_C  +  16 \Lambda^{2r+2},  \label{fCfQdef}
\end{equation}  with $r$ (again same as the rank of the gauge group) gauge invariant complex moduli $u_i$'s.
This can be easily obtained by taking no-flavor limit of Ref. \refcite{ArgyresShapere}.

 Observe in \eqref{fCfQdef} that $f_C=f_Q=0$ is possible only if $\Lambda=0$.  
In a quantum theory we demand $\Lambda\neq0$, so $f_C$ and $f_Q$ can never share a root. For any choices of moduli, $f_C$ and
$f_Q$ can never vanish at the same time. Just similarly to the $SU(r+1)$ case in Eqn. \eqref{sufactordisc}, the discriminant of the $Sp(2r)$ SW curve also factorizes as\cite{SD}
\begin{equation}\label{spfactordisc} \Delta_x f_{Sp(2r)}=(16\Lambda^{2r+2})^{2r} \Delta_x f_{C} \Delta_x f_{Q}.
\end{equation}
Again, in order for $f_{Sp(2r)}(x)$ to have a degenerate root, $f_C$ or $f_Q$ itself should have a degenerate root. 

We can study multiplicity of zeroes for $f_C$ and $f_Q$ separately without worrying about their roots getting mixed.
Again, just as in the $SU$ case, when we draw vanishing cycles and collision of branch points, we can use binary coloring. The branch points and vanishing cycles are all grouped into two mutually exclusive groups (for $C$ and $Q$ respectively.).  

  In order to give new names to two sets of branch points, let us introduce $C_i$'s and $Q_i$'s
as given in 
\begin{equation} 
f_C =  \prod_{i=1}^{r} (x-C_i), \qquad f_Q=  \prod_{i=0}^{r} (x-Q_i).
\end{equation}

 
   At discriminant loci $ \Delta_x f_{Sp(2r)}=0$, near the corresponding vanishing 1-cycle, the 1-form of \eqref{sprcurve1form} becomes infinitesimally small \cite{DSW}, far from becoming a delta function. This confirms that a singularity of the SW curve is indeed a singularity of the SW theory, again confirming the claim near Fig. \ref{local}.

 \section{Electric and Magnetic Charges of Massless Particles \label{monodromySec} }
 
So far, we discussed existence of massless particles in ${\cal N}=2$ theories. In this section, we will discuss electric and magnetic (dyonic) charges of these massless particles. Recall that the massless state was associated with a vanishing 1-cycle of Riemann surface, from Eqn. \eqref{tabledic} and Eqn. \eqref{Mlambda}. Dyonic charges can be read off by decomposing a 1-cycle into symplectic basis 1-cycles, as discussed near Eqn. \eqref{dyonchargecycle}. 

In the moduli space, there will be complex codimension 1 loci with a vanishing 1-cycle and it creates nontrivial topology on the moduli space with monodromy determined by the dyonic charge of the vanishing 1-cycle, as in Eqn. \eqref{PLformula}.

In this section, first we will look at dyon charges of the massless particles for the famous and simpler rank 1 case, and then move to higher rank cases reviewing the results obtained in Ref. \refcite{SD}, with focus on $SU(r+1)$ and $Sp(2r)$ case.

  \subsection{Rank 1 examples: how to read off monodromies of the Seiberg-Witten curves \label{rank1} }
 For pure SYM, the rank of gauge group $r$ equals to the genus of the SW curve and the number of complex moduli $u_i$'s, as one might recall from SW geometry of $SU(r+1)$ and $Sp(2r)$ gauge theory given in Eqns. \eqref{surcurve1form} and \eqref{sprcurve1form}. 
 Here we will warm-up by considering their 
  rank 1 cases, which 
  has one complex modulus $u$ for a genus-1 curve. Therefore the moduli space is a complex plane (real 2-dimensional surface), which we denote as $u$-plane. There exists non-trivial topology on the moduli space, created by existence of singular points on it. 
  
  By singular points on moduli space of SW curve, we mean the values of moduli which make the SW curve singular (i.e. with vanishing 1-cycles). Recall that it is equivalent to having zero discriminant of hyperelliptic curve. Demanding this single complex condition on the moduli space, we will have a complex codimension-1 loci in the moduli space as a solution set. 
  The modulus $u$ being the only parameter controlling the properties of a genus-one SW curve, vanishing discriminant condition will fix $u$ to possible isolated (separated) values. 
  
  First we will locate the singular points by discriminant condition, and then consider monodromy properties around each of them, by reading off dyon charges of vanishing 1-cycle, in the spirit of Eqn. \eqref{PLformula}. 
   Starting from a generic place in moduli space (a reference point $u_\ast$ on the moduli surface), we make non-contractible loops around each singular point (where discriminant vanishes), and consider monodromy along each path, associated with the singularity surrounded inside. 
 
Here we will discuss monodromy of rank $r=1$ cases of    
 $SU(r+1)$ SW curve given in Eqn. \eqref{surcurve1form} and
 $Sp(2r)$ SW curve given in  Eqn. \eqref{sprcurve1form}, which we will call $SU(2)$ and $Sp(2)$ curves. Since $SU(r+1)$ and $Sp(2r)$ gauge groups are identical at rank 1, these two distinct curves 
 in fact describe the same physical theories and indeed their monodromy properties match up with each other. Historically both curves were called $SU(2)$ SW curves, but we will call one of them $Sp(2)$ curve, because it has nice generalization for $Sp(2r)$ SW theories.
    
After absorbing some powers of two's into $\Lambda$ for convenience, $Sp(2)$ curve becomes
\begin{equation}
y^{2}=x\left( x(x-u)+\frac{1}{4}\Lambda ^{4}\right),   \label{sp2SW}
\end{equation} as first given in Ref. \refcite{SeibergWittenWithMatter}.
On the other hand, $SU(2)$
curve  \begin{equation}
y^{2}=(x^{2}-u)^{2}-\Lambda ^{4}=(x^{2}-{u+\Lambda ^{2}})(x^{2}-{u-\Lambda
^{2}}),  \label{su2Lerche}
\end{equation}   
follows from Eqn. \eqref{surcurve1form}, as first proposed by Ref.
 \refcite{KLYTsimpleADE}.
It is straightforward to check that discriminant vanishes at $u=\pm \Lambda^2$ for both SW curves for pure $SU(2)=Sp(2)$ theory given by \eqref{sp2SW} and \eqref{su2Lerche}.   
Now we will turn to finding out which 1-cycle of SW curve vanishes at $u=\pm \Lambda^2$.

 \subsubsection{Review of $SU(2)$ monodromy \label{su2monodromy}}

As explained in Ref. \refcite{KLYTsimpleADE}, $SU(2)$ curve has massless dyon and monopole at two different locations in the moduli space\footnote{In fact the moduli space has an extra singular point at $u =\infty$ as reviewed in Ref. \refcite{LercheReview}. 
However it is not associated with a particular vanishing 1-cycle, and its monodromy can be inferred from the knowledge of other singular points purely from consistency requirement. Therefore in this review we won't discuss the singular points at infinity in moduli space.}. 
The curve in Eqn. \eqref{su2Lerche} has four branch
points
\begin{equation}
 N_{0}= -N_{1} =   \sqrt{u+\Lambda^2}, \qquad P_{0}=  -P_{1}=  \sqrt{u-\Lambda^2}.
\end{equation}
At a generic value of modulus $u$, they are all distinct.
 As we vary $u$ toward two special values $u\rightarrow \pm \Lambda^2$, different pairs of branch points will collide: $N_0$ and $N_1$ collide as $u\rightarrow -\Lambda^2$ and $P_0$ and $P_1$ collide as $u\rightarrow  \Lambda^2$.
 
  \begin{figure}[htb]
\begin{center}
\includegraphics[
width=\textwidth 
]{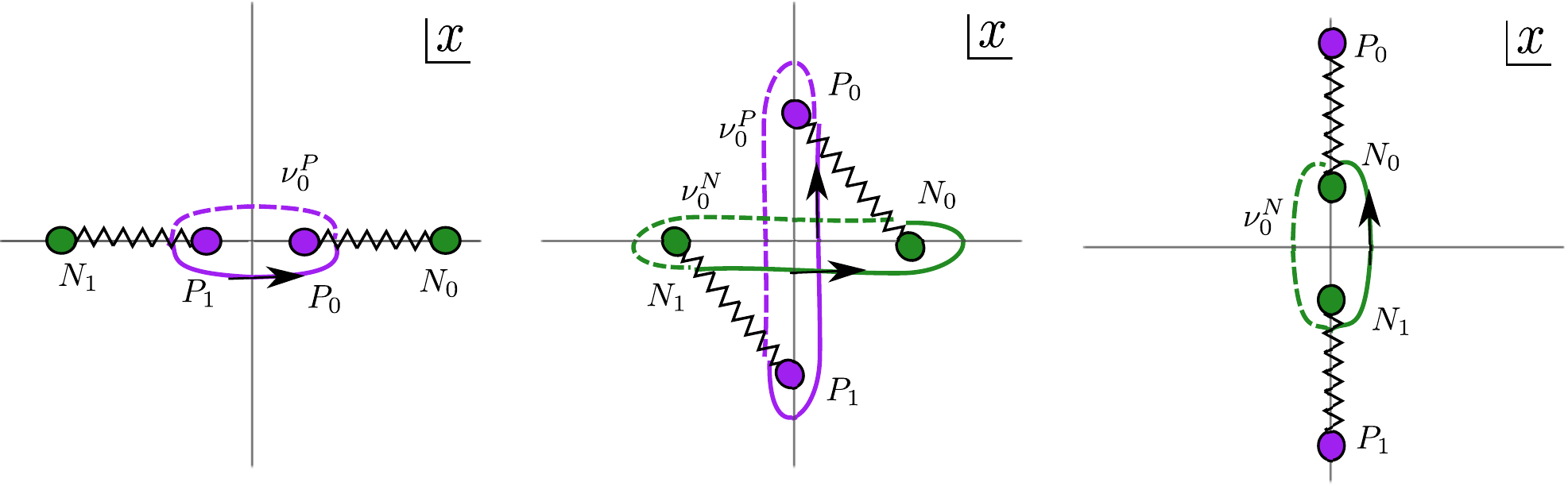}
\end{center}
\caption{Vanishing cycles for a pure $SU(2)$ Seiberg-Witten theory with its SW curve $y^{2}=(x^{2}-u)^{2}-\Lambda ^{4}=(x^{2}-{u+\Lambda ^{2}})(x^{2}-{u-\Lambda^{2}})$.
The branch points are drawn on the $x$-plane for varying values of the modulus $u$. From the left, $u$ takes the $u \sim \Lambda^2$, $u\sim 0$, $u\sim - \Lambda^2$ in the three figures drawn here.
 }
\label{rank1su}
\end{figure}

Fig. \ref{rank1su} denotes vanishing 1-cycles, associated to collision of those branch points.
The branch points are drawn on the $x$-plane for varying values of the modulus $u$. From the left, $u$ takes the $u \sim \Lambda^2$, $u\sim 0$, $u\sim - \Lambda^2$ in the three figures drawn here.
On the left of Fig. \ref{rank1su} ($u \sim \Lambda^2$), two purple branch-points $P_0$ and $P_1$ come close to each other. A 1-cycle drawn in purple denotes the corresponding vanishing 1-cycle, which goes through 2 branch cuts. Half of it is solid line, the other half is dashed line. Recalling that hyperelliptic curves are double-sheet fibration over an $x$-plane, we can take solid lines to be on the upper sheet ($y=\sqrt{f(x)}$) and dashed lines to be on the lower sheet ($y=-\sqrt{f(x)}$).
Each time a 1-cycle meets a branch-cut, it has to switch from solid to dash and vice versa.

Similarly, 
on the right of Fig. \ref{rank1su} ($u\sim - \Lambda^2$), two green branch-points $N_0$ and $N_1$ come close to each other. A 1-cycle drawn in green denotes the corresponding vanishing 1-cycle, which goes through 2 branch cuts. In the center of Fig. \ref{rank1su} ($u\sim 0$), all the branch points are separated, however it retains the topological information about 1-cycles which would vanish at $u = \pm \Lambda^2$.  

By counting their intersection number, we can read off from the figure that these two 1-cycles have mutual intersection number 2. 
We are allowed to choose an appropriate symplectic basis, so that we can they can be written as $\beta$ and $\beta-2\alpha$ ($\beta \circ (\beta-2\alpha)=-2$). In physics terms, we can perform electric-magnetic dualities to appoint them as a magnetic monopole and a dyon, in agreement with the usual convention. More details about these dyons are reviewed in Ref. \refcite{LercheReview} for example.
 
  \begin{figure}[htb]
\begin{center}
\includegraphics[
width=\textwidth 
]{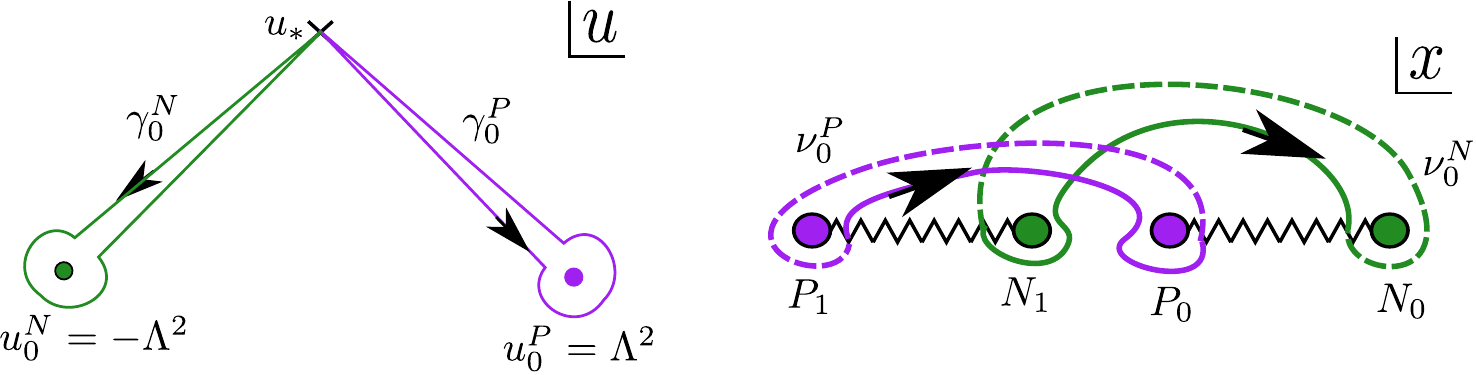}
\end{center}
\caption{Vanishing cycles for a pure $SU(2)$ Seiberg-Witten theory with its SW curve $y^{2}=(x^{2}-u)^{2}-\Lambda ^{4}=(x^{2}-{u+\Lambda ^{2}})(x^{2}-{u-\Lambda^{2}})$.
As $u$ varies on $u$-plane on green and purple paths given, the branch points will move along the path given in respective colors on $x$-plane. 
Note that the left side of the figure denotes the moduli space and the right side denotes the SW curve embedded in an ambient space whose coordinates are $x, y$. 
}
\label{su2}
\end{figure}

 The topological information of Fig. \ref{rank1su} is summarized in {Fig.
\ref{su2}}. On the left, moduli space is given. Green and purple paths denote the choice of how to vary the modulus $u$. As $u$ varies on the paths given, the branch points will move following the paths drawn on $x$-plane, as on the right of Fig. \ref{su2}. For higher ranks of $SU(r+1)$ SW theory in subsection \ref{SUdyon}, we will omit figures like Fig. \ref{rank1su} (which is a procedure how one obtains the information about vanishing 1-cycles) and only display figures similar to Fig. \ref{su2}, which contains full topological information of vanishing 1-cycles and trajectories in the moduli space.

Note that what trajectory each cycle takes does matter. It is important
not only which two branch points are connected, but also through what
trajectory they are connected. This should be clear from simple counting: with finite number of branch points ($N$), we can choose a finite number of pairs of branch points ($
\frac{N(N-1)}{2}$), however we have infinite (countable) number of distinct 1-cycles (as points on the $(N-2)$-dimensional lattice). 
 
As we will discuss later near Fig. \ref{abc} in subsection \ref{Sp4}, a different topological choice of trajectory on moduli space (on the left of Fig. \ref{su2}) translates to different dyon charges 
on the right. Therefore, we should associate each massless dyon not only with a singular loci on the moduli space, but also with topology of trajectory taken on the moduli space.

\subsubsection{$Sp(2)$ monodromy \label{sp1monodromy}}

Starting from the $Sp(2)$ curve of Eqn. \eqref{sp2SW}, shift $x$ by $x\rightarrow x+u$ to obtain
\begin{equation}
y^{2}=(x+u)\left( x(x+u)+\frac{1}{4}\Lambda ^{4}\right),   \label{sp2SWshift}
\end{equation}
 which identifies with expression given in Ref. \refcite{SeibergWittenWithMatter}.
Then perform $x\rightarrow 1/x, y\rightarrow x^{-2} y$ transformation to obtain
\begin{equation}
y^{2}=x(1+2 u x)\left( 1 +2 ux+ \Lambda ^{4} x^2 \right),   \label{sp2SWshift2}
\end{equation}
whose four branch points are
\begin{eqnarray}
O_{\infty}=0, \quad  C_1=-\frac{1}{2u}, \quad Q_{0 } &=&-\frac{1}{\Lambda^2}\left(u + \sqrt{-\Lambda^2+u^2}\right), \nonumber \\ Q_{ 1}&=&-\frac{1}{\Lambda^2}\left(u- \sqrt{-\Lambda^2+u^2}\right).\label{sp1fourpoints}
\end{eqnarray} 
 At generic value of $u$, all four branch points are separated, but as we vary $u$ into appropriate values, $Q_0$ and $Q_1$ collide with each other. 
  \begin{figure}[htb]
\begin{center}
\includegraphics[
width=\textwidth 
]{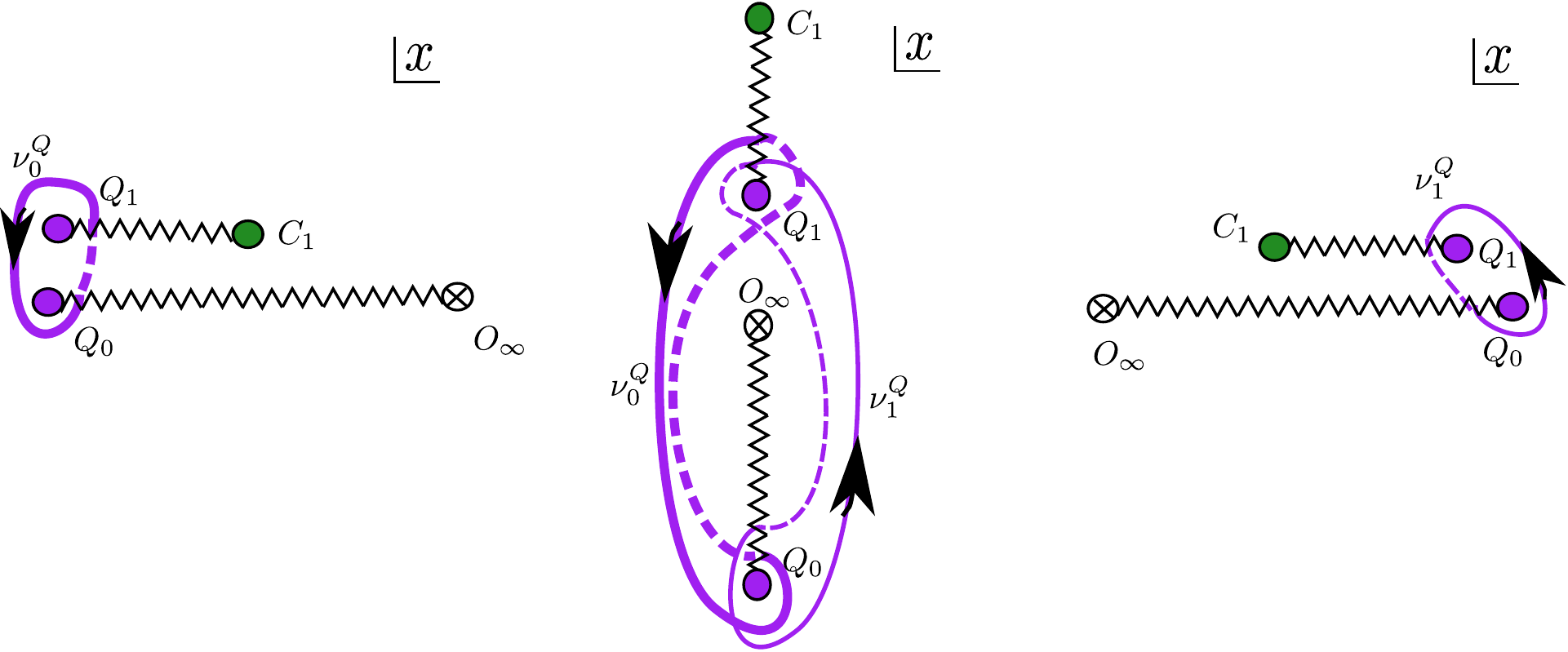}
\end{center}
\caption{Vanishing cycles for pure $Sp(2)$ Seiberg-Witten theory with a SW curve $y^{2}=x\left( x(x-u)+\frac{1}{4}\Lambda ^{4}\right)$. 
The branch points are drawn on the $x$-plane for varying values of the modulus $u$.}
\label{rank1spnoaxes2}
\end{figure}

 Fig. \ref{rank1spnoaxes2} shows a related animation. In the center figure, all branch-points are separated, and two purple 1-cycles (not vanishing here) are drawn for later convenience with two different thickness. On the left and right figures of Fig. \ref{rank1spnoaxes2}, $Q_0$ and $Q_1$ collide with each other. These figures preserve topological information about branch cuts, branch points, and 1-cycles. On the left, a thick 1-cycle vanishes, which is the same 1-cycle as in the center figure. On the right figure, a thin 1-cycle vanishes, and it is to be identified with the thin 1-cycle in the center figure. 
 
   \begin{figure}[htb]
\begin{center}
\includegraphics[
width=\textwidth 
]{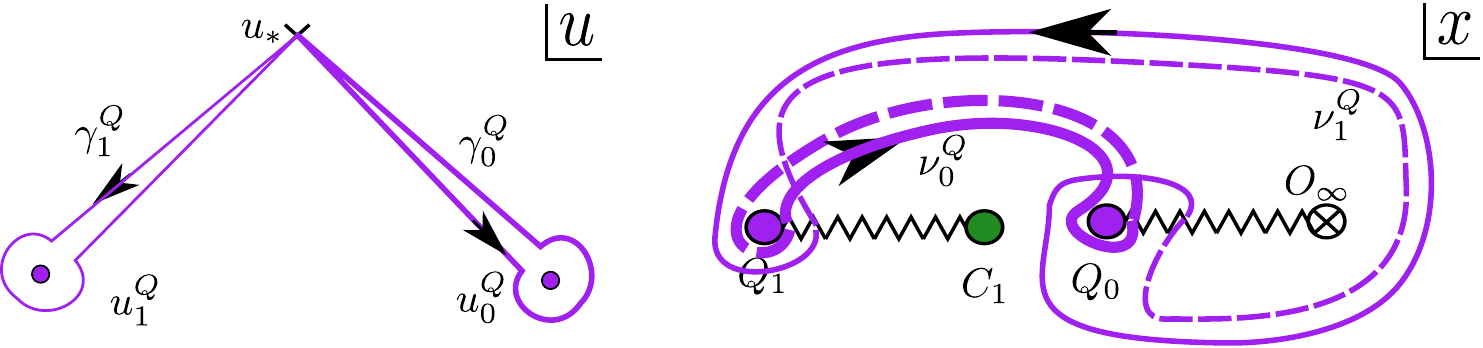}
\end{center}
\caption{Vanishing cycles for pure $Sp(2)$ Seiberg-Witten theory with a SW curve $y^{2}=x\left( x(x-u)+\frac{1}{4}\Lambda ^{4}\right)$. 
discriminant vanishes at $u=\pm \Lambda^2$. As $u$ varies on $u$-plane on thick and thin paths given, the branch points will move along the path in respective thickness on $x$-plane, 
forming vanishing 1-cycles.}
\label{sp1}
\end{figure}

Topological information of Fig. \ref{rank1spnoaxes2}  
is summarized in Fig. \ref{sp1} along with the choice of trajectory on the moduli space. 
On $u$-plane, there are two singularities, and depending on which singularity the path surrounds, $Q_0$ and $Q_1$ collide along different path. In Fig. \ref{sp1}, it is denoted with two different thickness of lines. 
 
 Again from the trajectories of branch points on the right of Fig. \ref{sp1}, we observe that the two 1-cycles have mutual intersection number 2. After an appropriate choice of symplectic basis, they can be identified as $\beta$ and $\beta-2\alpha$, or a magnetic monopole and dyon, matching the result for $SU(2)$ case (the same theory physically) above.

 \subsubsection{Moving on to the higher rank case}
   Now we move on to cases with higher and arbitrary rank $r$ of pure SYM with $SU(r+1)$ and $Sp(2r)$ gauge groups. Now we have $r$ complex moduli $u_1, \ldots, u_r$. Therefore the moduli space is a real $2r$-dimensional space. Discriminant loci are various real $(2r-2)$-dimensional loci embedded inside the moduli space. Again we employ the same technique as the rank 1 cases: starting from a generic place in moduli space, we make non-contractible loops around each component of discriminant loci, and read off dyon charges of vanishing 1-cycle associated. 

To render computation manageable, we take a real 2-dimensional slice of moduli space, on which discriminant loci are isolated points. We choose a generic point on that slice, and consider non-contractible loops, on that slice, around each singular point. Since we will be restricted to a modulus plane, the analysis will resemble that of rank 1 case, with the only difference being that we will have much more singular points on the modulus plane. Later in subsection \ref{Sp4}, we will consider taking multiple moduli slices, so that we can have multi-dimensional viewpoint on the singularity structure of the SW theory. 

For now, we consider monodromy of $SU(r+1)$ and $Sp(2r)$ SW theories confined on a certain moduli plane, which is chosen for the ease of computation.

\subsection{${\cal{N}}=2$ $SU(r+1)$ theory \label{SUdyon}}
    Here we discuss monodromy of the $SU(r+1)$ curve given in Eqn. \eqref{surcurve1form}. Again the moduli space has $r$ complex-valued coordinates $u_i$'s.
    Instead of considering the full real $2r$-dimensional moduli space, we consider its real 2-dimensional slice, a $u_{r-1}$-plane, after fixing all other $r-1$ moduli into constant values. We fix the first $r-2$ moduli to zero, and we also fix $u_r$ to a constant where its phase is chosen carefully as below 
      \begin{equation}
u_1= \cdots = u_{r-2}=0, \qquad  u_r/\Lambda^{r+1} \in i {\mathbb{R}}\label{uslice}
\end{equation} 
On such a hyperplane of moduli space, the discriminant simplifies into
\begin{equation}  \Delta_x f_{\pm} = (-1)^{\left[ \frac{r}{2} \right]}   r^{r}  (u_{r-1})^{r+1} +  (-1)^{\left[ \frac{r+1}{2} \right]} (r+1)^{r+1} (u_r  \pm \Lambda^{r+1})^r, \label{sudet}
\end{equation}
at whose vanishing loci the SW curve degenerates.

Note the symmetries on the moduli slice: there is $\mathbb{Z}_{r+1}$ symmetry from rotation of the phase on $u_{r-1}$. Also there is a $\mathbb{Z}_2$ symmetry for $\Delta_x f_{\pm}$, associated with complex conjugation of all $u_i$'s and switching between $\Delta_x f_{+}$ and $\Delta_x f_{-}$, due to fixed phase of $u_r$ as decided in Eqn. \eqref{uslice}. In Eqn. \eqref{sudet}, note that $u_r$ is a constant and only $u_{r-1}$ is a variable. Since $u_{r-1}$ has power $r+1$, there are $r+1$ solutions to each of $\Delta_x f_{+}=0$ and $\Delta_x f_{-}=0$. Those solutions have equal magnitude, and the phases are equally distanced among one another. 

\begin{figure}[htb]
\begin{center}
\includegraphics[
width=3.8in
]{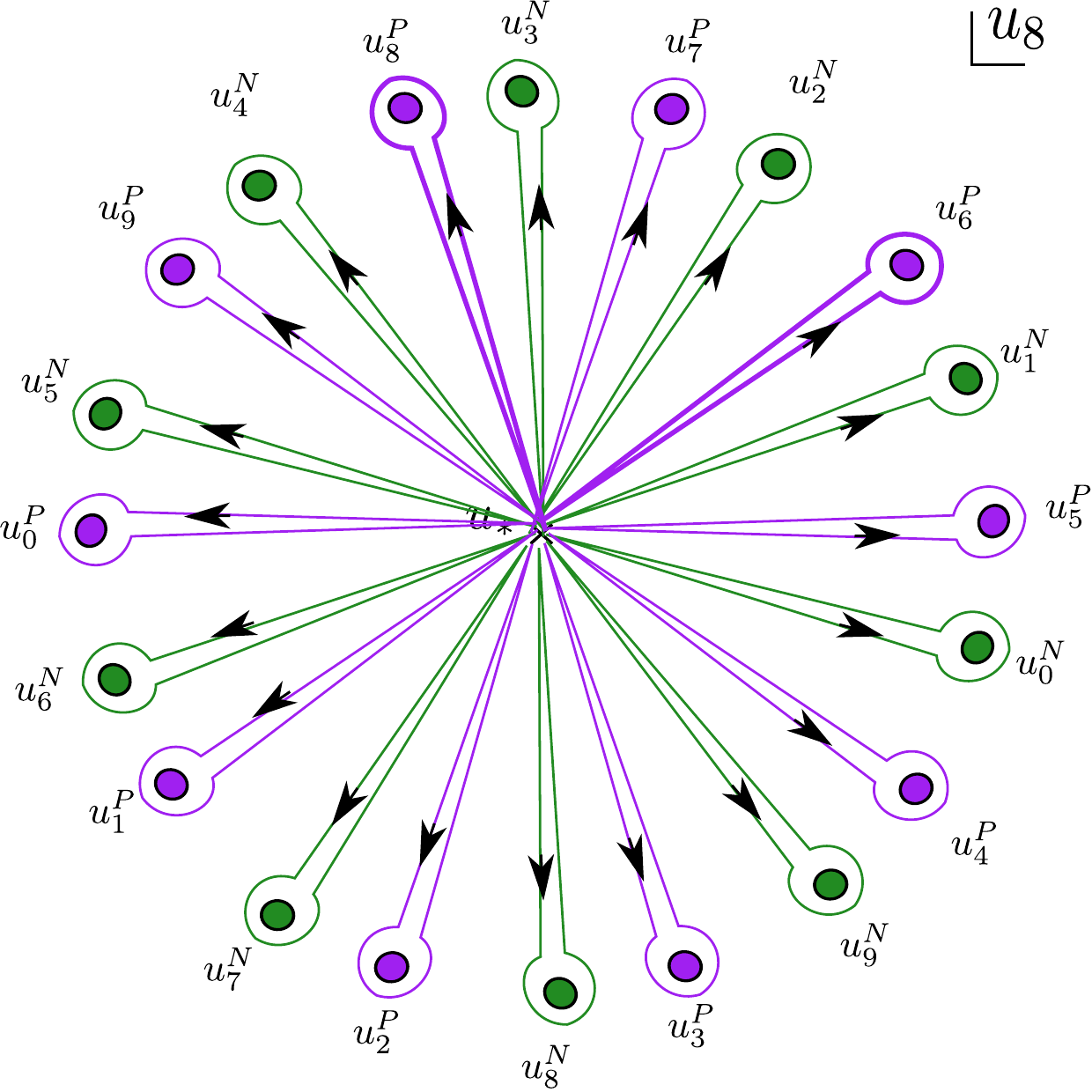}
\caption{ Singularity structure of $SU(10)$ curve on a moduli slice of the $u_8$-plane given by $u_1= u_2=\cdots = u_{r-2}=0, u_r = const$.
 Here $r=9$ case is drawn. 
 As we go around each singular point on the $u_8$-plane, we have a vanishing cycle on the $x$-plane as shown in {Fig.~\ref{rank9SU10xplane}}.}
\label{rank9SU10u8theta0}
\end{center}  
\end{figure}

If we recall from Eqn. \eqref{sufactordisc}, that $\Delta_x f \sim \Delta_x f_{+} \Delta_x f_{-}$, it follows $\Delta_x f \sim \Delta_x f_{+} \Delta_x f_{-}=0$ has $2r+2$ solutions on $u_{r-1}$-plane.
Therefore we can single out $2r+2$ points, arranged on a circle, on $u_{r-1}$-plane, where discriminant of the SW curve vanishes. This is depicted for rank 9 case in Fig. \ref{rank9SU10u8theta0} as eighteen marked points in purple and green. 
Each singular point on $u_{r-1}$-plane is responsible for vanishing of a 1-cycle and associated monodromy of Seiberg-Witten curve.

\begin{figure}[h]
\begin{center}
\includegraphics[
width=5in
]{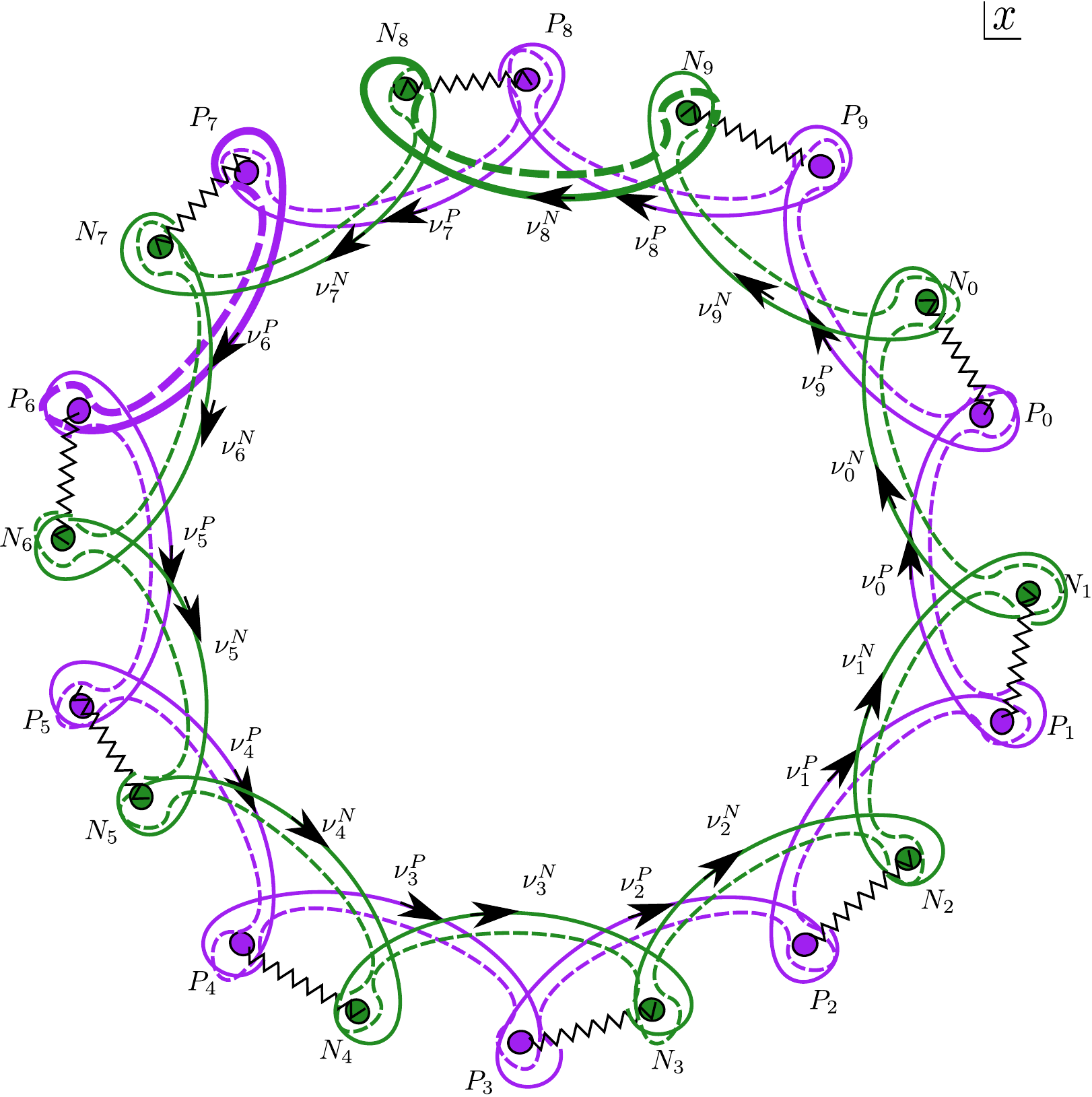} 
\caption{Drawn here are vanishing cycles for $SU(10)$ at a moduli slice, a $u_{8}$-plane given by $ u_1=u_2= \cdots = u_{7}=0$ and $\frac{ u_9}{\Lambda^{10} }\in i{\mathbb{R}} \cup \{ 0\}$.}
\label{rank9SU10xplane}
\end{center}  
\end{figure}

We pick a reference point at origin of $u_{r-1}$-plane, so that we manifest $\mathbb{Z}_{r+1}$ and $\mathbb{Z}_2$ symmetries on $x$-plane as well. When $u_{r-1}=0$, the branch points are spread on a circle on $x$-plane:
 $N_i$'s (and $P_i$'s) are even distributed on a circle with $\mathbb{Z}_{r+1}$ symmetry among themselves, and there is $\mathbb{Z}_2$ symmetry between $N_i$'s and $P_i$'s because of 
$\frac{ u_r}{\Lambda^{r+1} }$ is a pure imaginary number (or zero).
Rank 9 case is depicted in {Fig.~\ref{rank9SU10xplane}}: branch points on $x$-plane are drawn as green and purple marked points, at a reference point ($u_{r-1}=0$) on the moduli slice given by Eqn. \eqref{uslice}. 

As we vary $u_{r-1}$, approaching each of $2r+2$ singular points on $u_{r-1}$-plane, then a corresponding 1-cycle will vanish on $x$-plane, and we read off its dyon charge.
       On the $u_{r-1}$-plane, starting from the origin as the reference point, we make a non-contractible loop surrounding each singular point on $u_{r-1}$-plane (as shown in Fig. \ref{rank9SU10u8theta0}).
 As we vary $u_{r-1}$, we observe the vanishing cycles on the 
$x$-plane, as in {Fig.~\ref{rank9SU10xplane}} for rank 9 case:
the branch cuts are drawn connecting the branch points of corresponding colors.

\begin{figure}[h]
\begin{center}
\includegraphics[
width=\textwidth]{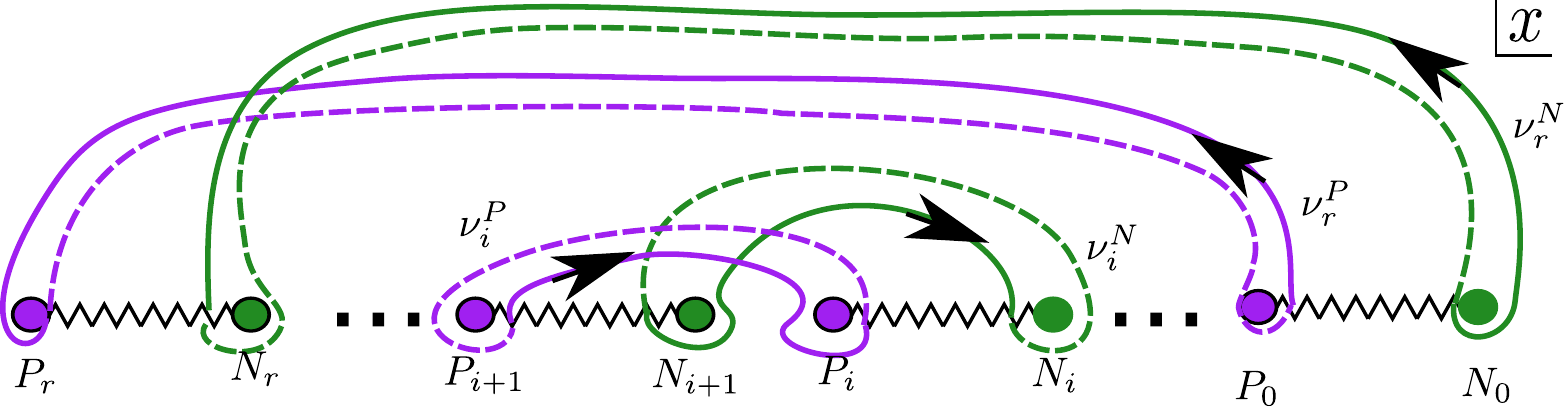}
\caption{Vanishing 1-cycles for $SU(r+1)$ curve}
\label{ardyon}
\end{center}  
\end{figure}

Also, Fig. \ref{ardyon} (we stretched out the $x$-plane, opening up the circle on the $x$-plane into a line) is given for general ranks. Vanishing 1-cycles satisfy
\begin{equation}
 {\color{purple}\nu_{i}^P}  \circ  {\color{purple}  \nu_{i+1}^P}=   {\color{forestgreen}\nu_{i}^N} \circ    {\color{forestgreen}\nu_{i+1}^N}= -1 , \quad
 {\color{purple} \nu_{i}^P}  \circ     {\color{forestgreen}\nu_{i}^N}=  2 , \quad   {\color{purple}\nu_{i}^P}  \circ     {\color{forestgreen}\nu_{i+1}^N}=- 2,
\end{equation}
while all other intersection numbers vanish.

 \begin{figure}[h]
\begin{center}
\includegraphics[
width=\textwidth]{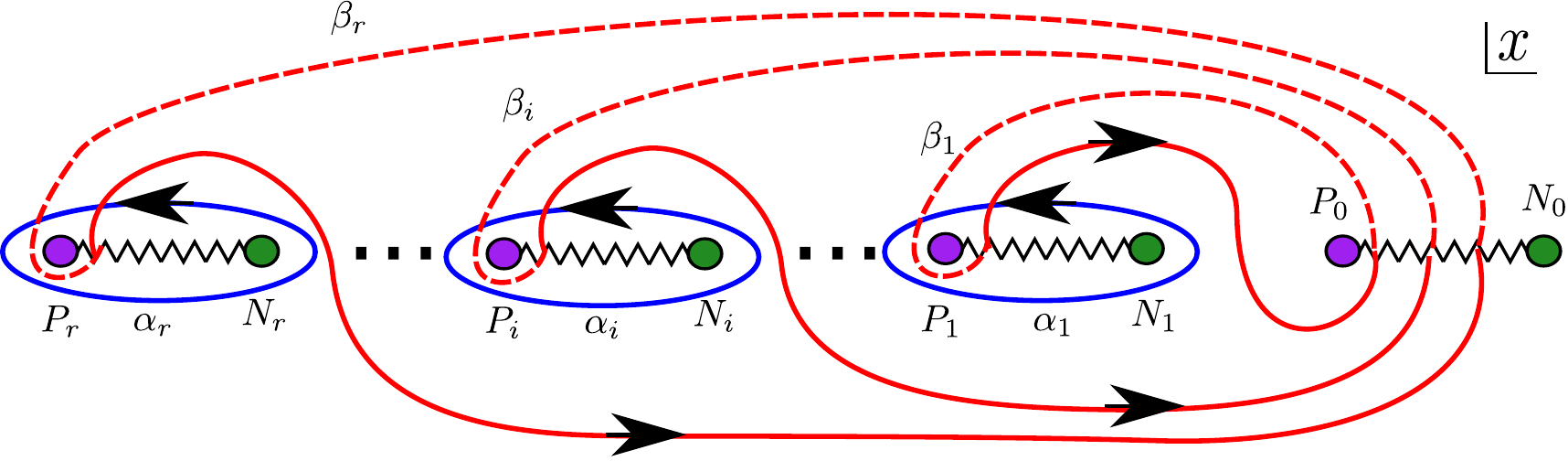}
\caption{A particular choice of symplectic basis cycles for $SU(r+1)$ curve}
\label{arcyclesbasis}
\end{center}  
\end{figure}
   
 To read off the dyon charges of massless states, we choose a set of symplectic bases. 
All choice is equivalent to up to electromagnetic duality.
With a symplectic basis given in {Fig.~\ref{arcyclesbasis}}, we are choosing $\alpha_i$ cycles to go around each branch cut connecting $P_i$ and $N_i$ branch points. 
We are choosing $\beta_i$ cycles to connect between $P_0$ and $P_r$ branch points. 
In the {Fig.~\ref{arcyclesbasis}},
to make it convenient to generalize to arbitrary ranks, we rearranged the branch cuts on the $x$-plane into a line. 
With such a choice of symplectic bases, we can write down the vanishing 1-cycles as:
\begin{eqnarray}    
 {\color{purple}\nu_{0}^{P}}&=&\beta_{1}, \qquad      {\color{forestgreen}\nu_{0}^{N}}=\beta_{1}+\sum_{i=1}^{r}\alpha_{i}+\alpha_{1}, \qquad   {\color{purple} \nu_{r}^{P}}=-\beta_{r}+\sum_{i=1}^{r-1}\alpha_{i}, \qquad  {\color{forestgreen}  \nu_{r}^{N}}=-\beta_{r}-2\alpha_{r},  \nonumber   \\
 {\color{purple}\nu_{i}^{P}} &=&\beta_{i+1}-\beta_{i}-\alpha_{i}, \qquad  
 {\color{forestgreen}\nu_{i}^{N}}=\beta_{i+1}-\beta_{i}+\alpha_{i+1}-2\alpha_{i}, \qquad i=1,\cdots,r-1. \label{sucharges}  
\end{eqnarray}

We will end this subsection by how this analysis (mostly based on Ref. 
\refcite{SD}) fits in the existing literature.
  Recall that 
 Seiberg-Witten curves and massless dyon charges were much-studied for low rank cases. 
Original Seiberg-Witten paper\cite{SeibergWittenNoMatter, SeibergWittenWithMatter} 
studied the  
curves, massless dyons (monodromies), and some aspects of singularity aspects for 
$SU(2)$ theory with and without matter. Refs. \refcite{KLYTsimpleADE,KLYTmndrmSU} have discussed 
 six vanishing cycles of the SW curve for pure $SU(3)$ on a slightly different moduli slice from the one chosen here. 
Vanishing 1-cycles of $SU(n)$ SW curve were also studied in Refs. \refcite{Fraser:1996pw,Chen:2011gk}.
There are lots of recent developments in obtaining BPS spectra (including massive ones) as discussed in Refs.
\refcite{cordovafa0,Gaiotto:2012db,LonghiGen,Chuang:2013wt} for example.

\subsection{Monodromies of pure ${\mathcal N}=2$ $Sp(2r)$ theories \label{spSec}}

In the last subsection we discussed pure $SU(r+1)$ SW theories and computed their monodromies. Here we
will continue the similar analysis for $Sp(2r)$ SW theories. Seiberg-Witten curves for pure $Sp(2r)$ theory is given by\cite{SD}
\begin{equation}
y^2 =  x \left(1+\sum_{i=1}^r u_i x^i \right)\left(1+\sum_{i=1}^r u_i x^i +x^{r+1} \right) \label{spcurve}
\end{equation}
after appropriate coordinate transformations of no-flavor limit of Ref. \refcite{ArgyresShapere} and setting $\Lambda \ne 0$ to satisfy $16 \Lambda^{2r+2}=1$ without loss of generality.

 \begin{figure}[h]
   \begin{center}
        \includegraphics[width=\textwidth]{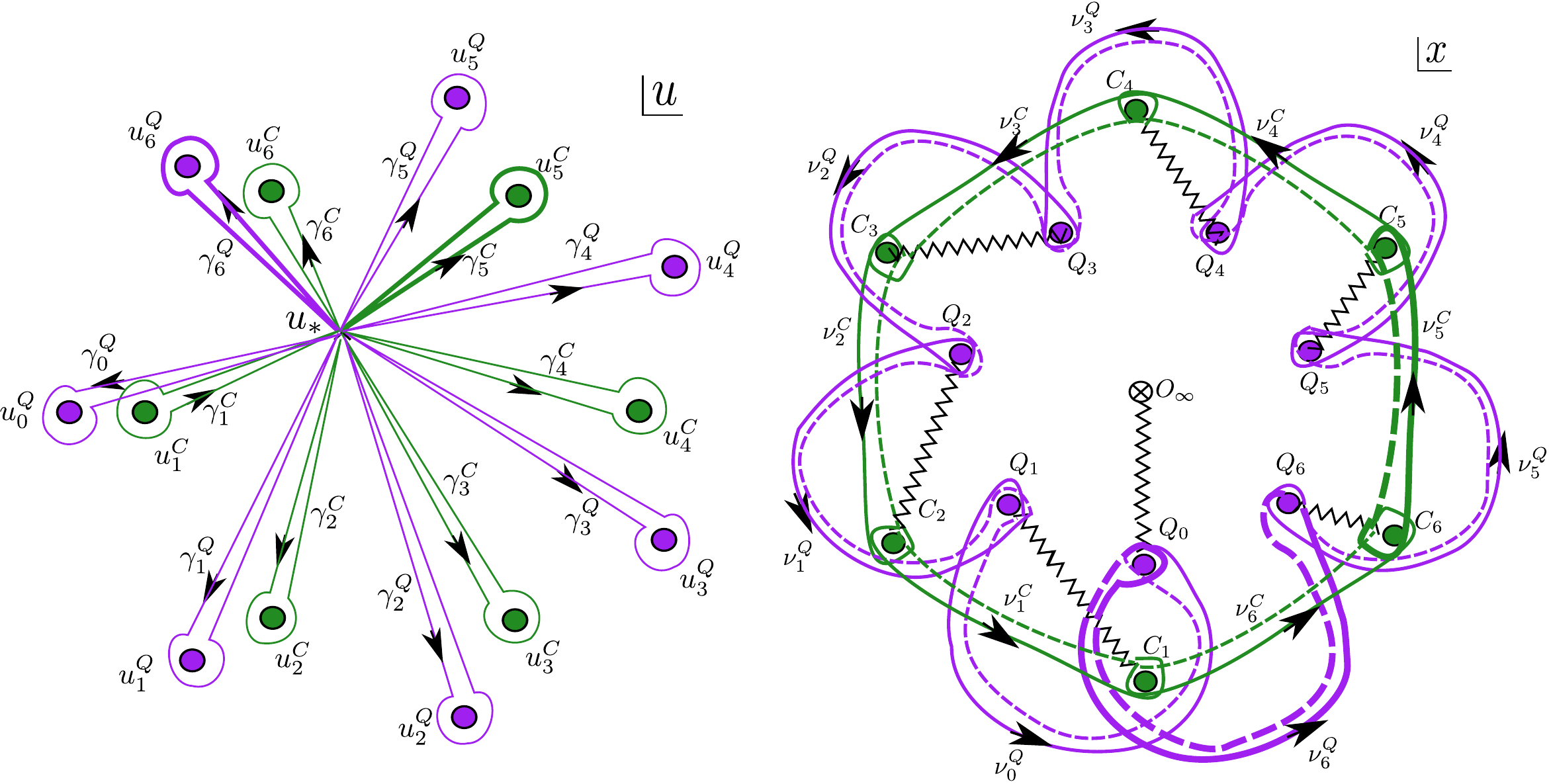}
    \end{center}
    \caption{Vanishing cycles of $Sp(12)$ curve at a moduli slice, which is given by a $u_1$-plane of $u_2=\cdots=u_{r-1}=0, u_r=1/9$}
    \label{sp6flower}
   \end{figure} 

Vanishing cycles are computed in Ref. \refcite{SD} in a certain moduli slice for $r \le 6$ by plotting in Mathematica, and similar form is conjectured for higher ranks. Here we summarize the result.

We choose a moduli hyperplane to be a $u_1$-plane given by fixing $u_2=\cdots=u_{r-1}=0$ and setting $u_r$ to be a fixed small number. Choose the origin $u_1=0$ as a reference point. 
Up to rank 6, if we choose $u_r$ to be small enough
then branch points on the $x$-plane are arranged such that all the $Q_i$'s are surrounding origin $O_\infty$ (a branch point at $x=0$ in Eqn. \ref{spcurve}), and all the $C_i$'s are surrounding all the $Q$ points. 
Vanishing 1-cycles have the following non-zero intersection numbers:
\begin{eqnarray}\nu_{i}^Q\circ \nu_{i+1}^Q &=& \nu_{i}^C \circ \nu_{i+1}^C= 1,\quad \nu_{i}^Q\circ \nu_{i}^C= \nu_{i}^C \circ \nu_{i-1}^Q=2, \qquad i = 1, \ldots, r 
,\nonumber \\
\nu_{r}^Q\circ \nu_{0}^Q&=& 3, \quad \nu_{0}^Q\circ \nu_{r}^C=\nu_{1}^C \circ \nu_{r}^Q=-2.  \label{spIntNum}
  \end{eqnarray}
Rank 6 case is depicted in {Fig. \ref{sp6flower}}. Unlike the $SU(r+1)$ case, 
it is difficult to find an exact method to obtain the vanishing cycles for arbitrary high ranks of $Sp(2r)$. 
Instead, we compute the vanishing cycles in some patches of the moduli space for low ranks, and read off a pattern to conjecture for general ranks. 
Ref.
\refcite{SD} conjectures that it is always possible to choose $u_r$ to be small enough such that all the $Q_i$ points are inside the $\nu^C$ cycles, such that \eqref{spIntNum} holds, for any rank $r$.  

   We write down these $2r+1$ vanishing cycles 
    \begin{eqnarray}  
   \nu_{i}^{C}=-\beta_{i}+\beta_{i+1}+\alpha_{i+1} , \quad
 \nu_{i}^{Q}=   \nu_{i}^{C}  -\alpha_{i}+ \alpha_{i+1},     \quad
   i=1,\cdots,r-1  , \nonumber \\
 \nu_{r}^{C}=\beta_{1} -\beta_{r} -\sum_{i=2}^{r}\alpha_{i}, \quad
 \nu_{0}^{Q}=\beta_{1} + 2\alpha_1, \quad  
 \nu_{r}^{Q}= \nu_{r}^{C}+ \beta_1  + \alpha_1 -\alpha_{r}. \label{spcharges}
  \end{eqnarray}
in terms of symplectic basis given in {Fig.~\ref{sprcyclesbasis}}. 
We are choosing $\alpha_i$ cycles to go around each branch cut connecting $Q_i$ and $C_i$ branch points. We are choosing $\beta_i$ cycles to connect between $Q_0$ and $Q_r$ branch points.  

\begin{figure}[pb]
\begin{center}
\includegraphics[
width=\textwidth
]{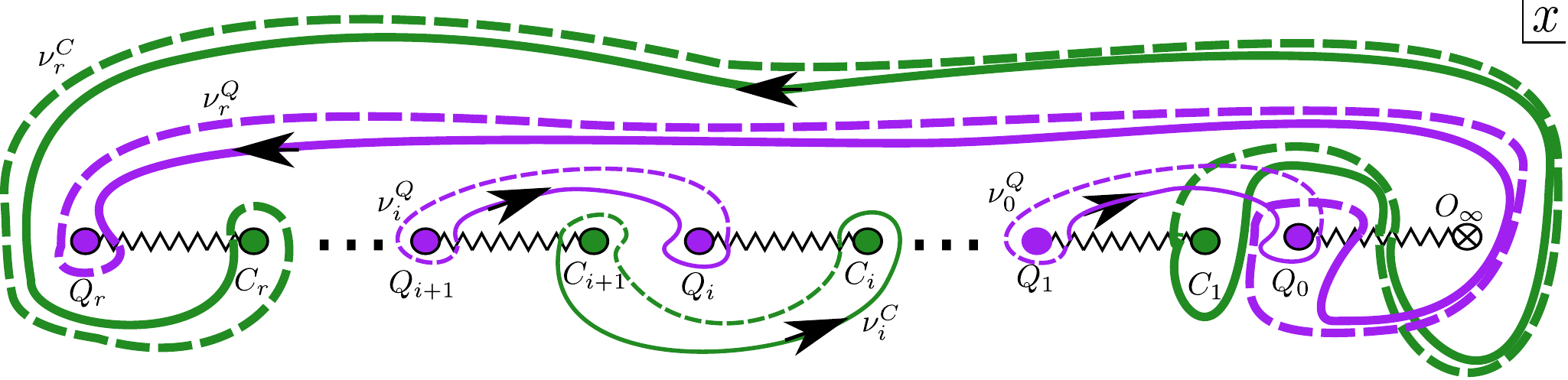}
\end{center}
\caption{Vanishing 1-cycles for $Sp(2r)$ curve}
\label{sprdyon}
\end{figure}

\begin{figure}[pb]
\begin{center}
\includegraphics[
width=\textwidth
]{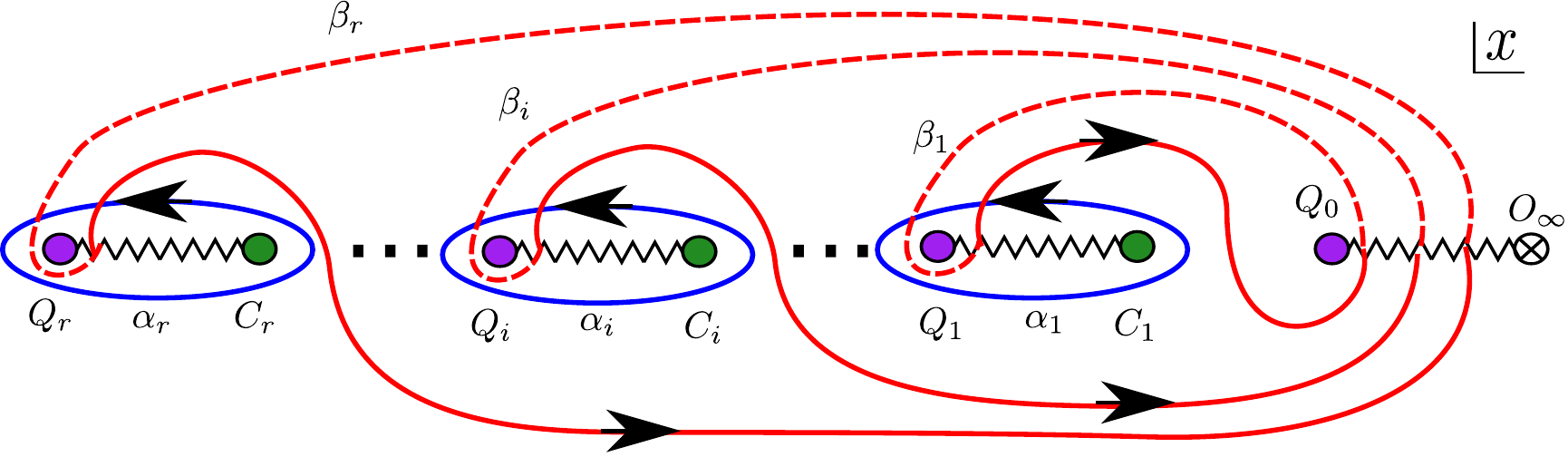}
\end{center}
\caption{A particular choice of symplectic basis cycles for the $Sp(2r)$ curve}
\label{sprcyclesbasis}
\end{figure}

\section{Argyres-Douglas Loci: Massless Electron \& Massless Magnetic Monopole  \label{ArDo}}
In previous section we studied vanishing 1-cycles at a complex codimension 1 loci of the moduli space. 
Demanding one 1-cycle to vanish takes away 1 complex degree of freedom, and therefore such loci have one dimension less than the full moduli space.
In this section, we consider singularity loci of moduli space with codimension 2 (or more), where we demand 2 (or more) 1-cycles to vanish. 
 This section will discuss degeneration of Seiberg-Witten curves so that mutually intersecting 1-cycles vanish at the same time. 
Such geometry (reviewed in subsection \ref{ReviewN2}) corresponds to Argyres-Douglas theories, containing massless dyons which are mutually non-local 
(they cannot turn into pure electric particles by any electromagnetic dualities) as studied in Ref. \refcite{ArgyresDouglas}.

In this review, our first encounter with Argyres-Douglas theory will be in the context of $Sp(4)$ SW theory. Now we take a closer look at various singularity loci with complex-codimensions 1 and 2 inside the moduli space.

\subsection{Singularity structure of $Sp(4)=C_2$: detailed look on BPS spectra \label{Sp4}}
 
Similarly to the previous section, we compute the dyon charges of vanishing 1-cycles for a SW curve, for $Sp(4)$ gauge group. 
Instead of confining ourselves on a moduli slice, we will consider a set of moduli slices. We will observe how vanishing cycles change as we change the choice of hypersurface.

\begin{figure}[htb]
\begin{center}
\includegraphics[width=\textwidth]{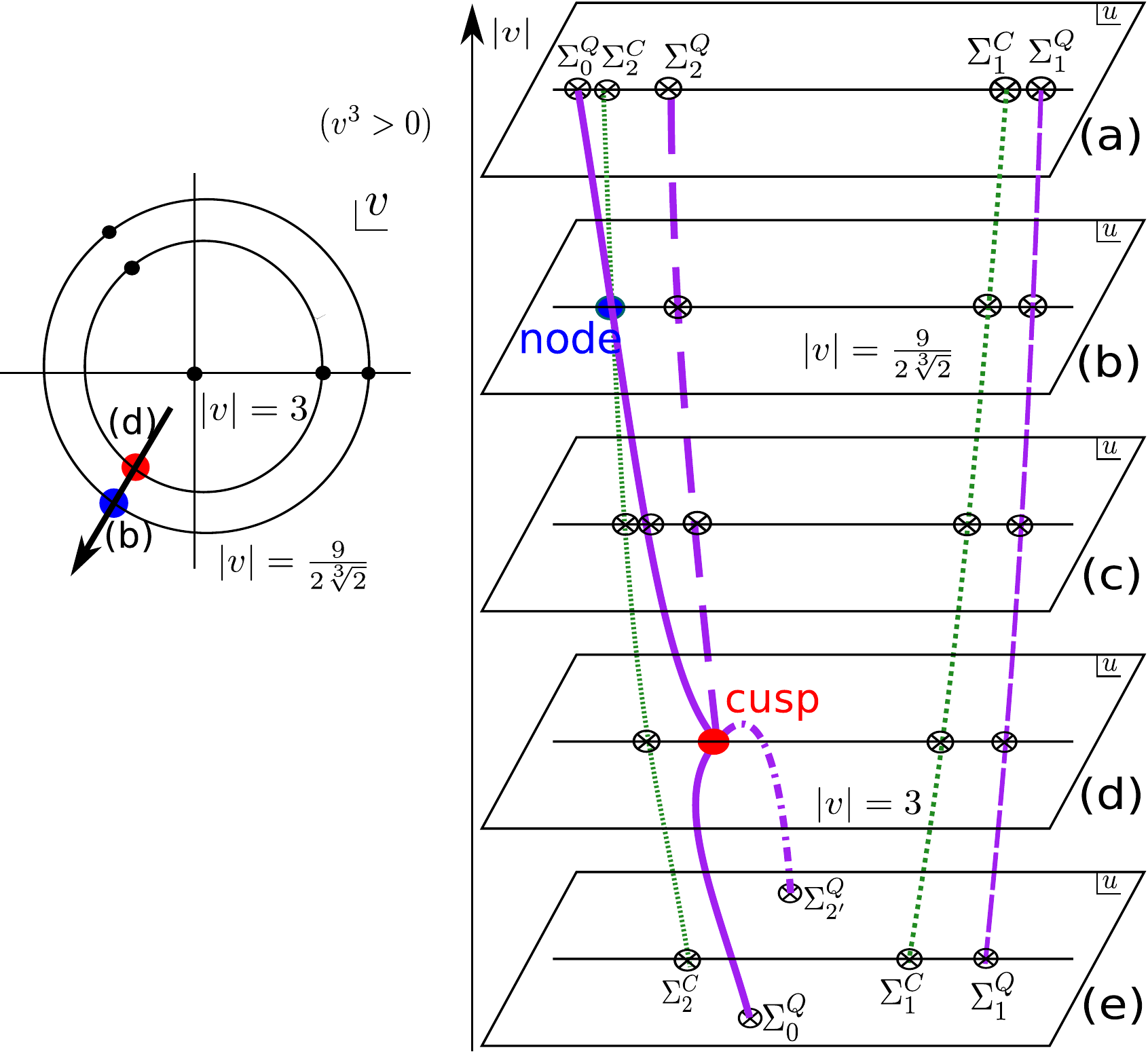}
\end{center}
\caption{A subspace inside the moduli space of pure $Sp(4)$ Seiberg-Witten theory.
Vanishing discriminant loci $\Sigma_2^Q$ and $\Sigma_{2^\prime}^Q$ are drawn with two different types (big and small) of dashed lines. They get interchanged at a cusp point (red).}
\label{moduli3dcolor}
\end{figure}

 Since the gauge group has rank 2,
 $Sp(4)$ SW theory has a 2 complex (4 real) dimensional moduli space whose coordinates are two complex moduli $u\equiv u_1, v\equiv u_2$.
Instead of full 4 real dimensional moduli space, we will take a $3$ real dimensional subspace by fixing the phase of $v$ such that $v^3$ is real, as in the left side of
{Fig.~\ref{moduli3dcolor}}. 
This choice was made so that the moduli subspace contains all interesting singular points inside, where multiple 1-cycles vanish at the same time\footnote{As we will discuss later, all solutions to the vanishing double discriminant given in Eqn. \eqref{doublediscsp4} satisfy that $v^3$ is real. This is a necessary, but not sufficient, condition for two 1-cycles to vanish.}. 

As we vary magnitude of $v$, we consider singularity structure on each $u$-plane: we locate where the curve degenerates, and we compute the corresponding dyon charge for vanishing 1-cycle.

On the right of {Fig.~\ref{moduli3dcolor}}, each of the five $u$-planes, marked with (a) to (e), is a slice of the moduli space at
different magnitude of $v$. The purple and green curves running vertically are denoted by $\Sigma^{C}_i$'s and $\Sigma^{Q}_i$'s. These are singularity loci in the moduli space with complex codimension 1. This is where the corresponding 1-cycle vanishes (i.e. a dyon becomes massless), and it is captured by the vanishing
discriminant of the curve (one complex condition). 

Interesting things happen when discriminant loci $\Sigma$'s intersect inside moduli space, at complex codimension 2 loci, forming a worse singularity. 
There, two 1-cycles vanish at the same time. In other words, massless dyons coexist. For example, at a blue point labelled `node' in the right of {Fig.~\ref{moduli3dcolor}}, $\Sigma^{Q}_0$ and $\Sigma^{C}_2$ intersect, and that is where 1-cycles denoted by 
$\nu^{Q}_0$ and $\nu^{C}_2$ degenerate at the same time. 

When two 1-cycles vanish at the same time, the SW curve (embedded in the ambient space whose coordinates are $x,y$) degenerates into either cusp or node form. 
The shape of intersection loci of $\Sigma$'s also takes cusp or node form respectively inside the moduli space. Each leads to different
kind of singularity, mutually non-local and local massless dyons. This is not tied to $Sp(4)$ gauge group, and similar phenomena occur for pure $SU(3)$ theory, too\cite{KLYTsimpleADE}.

When $\Sigma$'s (discriminant loci) intersect each other, it is seen as colliding of singular points on the corresponding $u$-plane, the moduli slice. 
For example, on the three $u$-planes marked by (a), (c), and (e), there are five singular points where $\Sigma$'s pierce through. 
On the two $u$-planes marked by (b) and (d), two of the singular points are on top of each other, so we see only 4 separate points on $u$-plane. 
 
Recall from subsection \ref{GeomReview} that collision of branch points on the $x$-plane was captured by vanishing of discriminant operator with respect to $x$, $\Delta_x$ acting on $f$.
Similarly, when the singular points collide on the $u$-plane, it is captured by another discriminant operator with respect to $u$, $\Delta_u$ acting on $\Delta_x f$. 
In other words demanding vanishing discriminant $  \Delta_x f =0 $ and double discriminant $ \Delta_u \Delta_x f =0 $ bring out all the candidates for having two vanishing 1-cycles. 
However among this, only those which satisfy $ d \Delta_x f =0 $ truly qualifies as a singularity loci of discriminant loci of hyperelliptic curve. The points where $ \Delta_u \Delta_x f =\Delta_x f =0 $
and $ d \Delta_x f \ne 0 $ is where the discriminant loci is smooth but it appears singular on a particular slice of moduli space. We will elaborate more near {Fig.~\ref{figure8}} later. 

 In the case of $Sp(4)$ SW theory, vanishing double discriminant gives one constraint as below:
\begin{eqnarray}   
 \Delta_u \Delta_x f_{Sp(4)}&=& 2^8 v (v^3-3^3)^3 (2^4 v^3- 3^6)^2 , 
 \label{doublediscsp4}
   \end{eqnarray}    
   whose roots 
   \begin{equation}
v=\Bigg\{ 0, 3 \alpha_3^i, \frac{9}{2 \sqrt[3]{2}} \alpha_3^j \Bigg\}
   \end{equation}
   are marked by seven dots in the left of the {Fig.~\ref{moduli3dcolor}}. 
   One of the solutions of Eqn. \eqref{doublediscsp4} is $v=0$, but it does not translate into having two massless dyons.      
      The other six points
 correspond to having two massless BPS dyons. In subsection \ref{exteriordVSdd} we will discuss how $d\Delta_x f=\Delta_x f=0$ is equivalent to having two massless BPS dyons. 
$v=0$ does not satisfy that relation, but the rest 6 does (with proper choice of $u$ value). 

Note the degeneracy of roots to the double discriminant of Eqn. \eqref{doublediscsp4}. The roots $3 \alpha_3^i$ and $ \frac{9}{2 \sqrt[3]{2}} \alpha_3^j $ have degeneracy 3 and 2 respectively.  
In the next section we will review that this is a universal criterion for having mutually non-local and local massless dyons.  
 Actually vanishing of double discriminant is a necessary condition for having multiple massless dyons, but not a sufficient condition as we will see in subsection \ref{exteriordVSdd}.

\begin{figure}[pb]
        \begin{center}
\includegraphics[height=19.3cm]{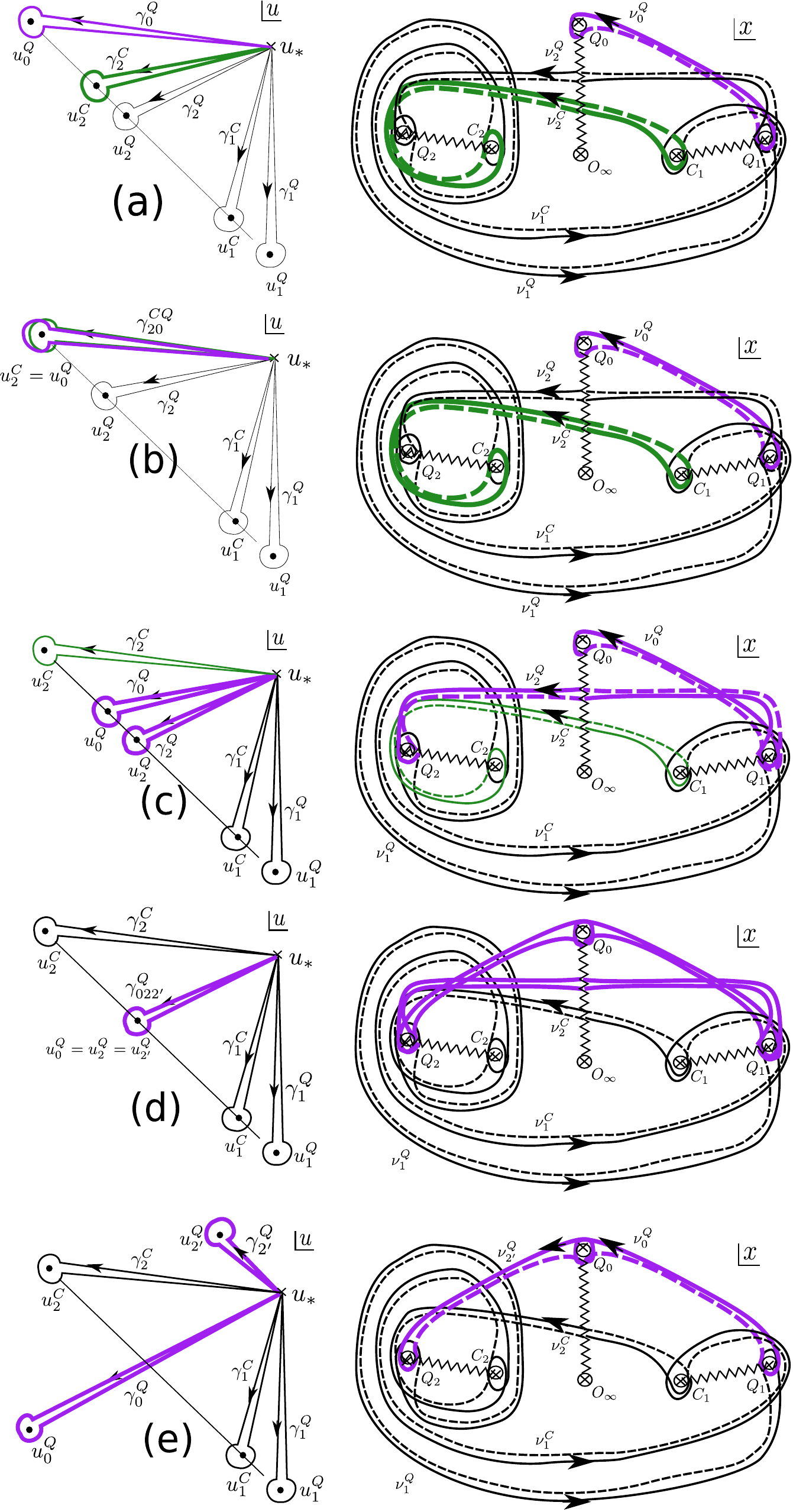}
        \end{center}
        \caption{Singularity structures at slices of the moduli space for pure $Sp(4)$ Seiberg-Witten theory.}\label{c2_10pic_down}
        \end{figure}

For each of $u$-planes marked by (a) to (e) of {Fig.~\ref{moduli3dcolor}}, we
have drawn a corresponding $x$-plane in {Fig.~\ref{c2_10pic_down}} displaying the vanishing cycles
on the $x$-plane, for each moduli slice.  
Let us have a closer look staring from the
top slice marked as (a).

\begin{description}
\item[(a)] choose a $u$-plane where $|v| > \frac{9}{2 \sqrt[3]{2}}$, then 
vanishing discriminant loci of the SW curve (marked as $\Sigma^{C}_{i}$'s and $\Sigma^{Q}_{i}$'s in {Fig.~\ref{moduli3dcolor}}) intersect the $u$-plane at five separate points $u^{C}_{i}$'s and $u^{Q}_{i}$'s. 
Each of the 5 singularity points on the $u$-plane ($u^{C}_{i}$'s and $u^{Q}_{i}$'s) is responsible for one vanishing 1-cycle ($\nu^{C}_{i}$'s and $\nu^{Q}_{i}$'s with corresponding choice of sub-/super- scripts). 
By observing the relative trajectory of branch points on the 
$x$-plane (with help of plotting in Mathematica), we can read
off the 5 vanishing cycles as below:
\begin{eqnarray} 
\nu_{0}^{Q}&=&\beta_{1} , \quad
  \nu_{1}^{Q}=-\beta_1 + \beta_{2} +\alpha_{1} +4 \alpha_2    , \quad  
  \nu_{2}^{Q}=-\beta_1-\beta_{2}-\alpha_{1}-2\alpha_{2}, \nonumber \\
  \nu_{1}^{C}&=&-\beta_1 + \beta_{2} +2\alpha_{1} +3 \alpha_2  , \quad 
    \nu_{2}^{C}=-\beta_1 - \beta_{2} - \alpha_2 .  \label{cycle}
  \end{eqnarray}

\item[(b)] As we change the moduli $|v|$ so that  $|v| = \frac{9}{2 \sqrt[3]{2}}$, singularity points on the $u$-plane, $u^Q_0$ and $u^C_2$, now collide.
 Corresponding 1-cycles $\nu^Q_0$ and $\nu^C_2$ vanish simultaneously at the point given by $u=u^Q_0 = u^C_2$ and $|v| = \frac{9}{2 \sqrt[3]{2}}$.   
However these two cycles are mutually local as seen from Eqn. \eqref{cycle} or Fig. \ref{c2_10pic_down}: responsible four branch points on the $x$-plane collide pairwise ($C_1 \leftrightarrow C_2$, $Q_0 \leftrightarrow Q_1$). 
The SW curve degenerates into a node form, $y^2 \sim (x-C)^2 (x-Q)^2 \times \cdots$. 
Singularity loci (of vanishing $\Delta_x f $) $\Sigma_{0}^{Q}$ and $\Sigma_{2}^{C}$ intersect at $|v|=\frac{9}{2\sqrt[3]{3}}$, with node-like
crossing. Note that the node-like singularity appears both in the ambient space and the moduli space of SW curve.

\item[(c)] We can change the moduli $|v|$ further to enter the range  
 $3 < |v| < \frac{9}{2 \sqrt[3]{2}}$.
Unlike (b), $u^Q_0$ and $u^C_2$ are separated again restoring back to (a). Dyon charges of the vanishing cycles  for (a), (b), and (c) stay the same as given by Eqn. \eqref{cycle}.  

\item[(d)] As we change the magnitude of the moduli $v$, to satisfy  $ |v|=3$, discriminant loci $\Sigma _{0}^{Q}$ and $\Sigma _{2}^{Q}$ intersect on the $u$-plane, forming a cusp-like singularity inside the moduli space. 
Two singular points $u_{0}^{Q}$ and $u_{2}^{Q}$ collide on the $u$-plane, and the vanishing cycles $\nu_{0}^{Q}$ and $\nu_{2}^{Q}$ {\it merge} into something which is no longer a 1-cycle. Now it is not possible to tell apart vanishing 1-cycles\footnote{Before we identified a vanishing 1-cycle by collision of 2 branch points. Their trajectory formed a 1-cycle, which goes through 2 branch cuts. Half of it was solid line, the other half was dashed line, in figures here. Recall that each time a 1-cycle meets a branch-cut, it has to switch from solid to dash and vice versa. Now we have 3 branch point colliding together. Now we the trajectory has 3 pieces involving 3 branch cuts, and there is no consistent way of assigning dash and solid for the odd number of pieces.}. 
Now three branch points $Q_{0}, Q_1,$ and $Q_2$ collide at the same time on the $x$-plane. 
Two cycles $\nu_{0}^{Q}$ and $\nu_{2}^{Q}$ become massless at the same time, but they are mutually {\it non-local} as clear from Eqn. \eqref{cycle}. The SW curve
degenerates into a cusp form $y^{2}\sim (x-a)^{3}\times \cdots $, giving
Argyres-Douglas theory, with two mutually non-local massless BPS dyons. Note that the cusp-like singularity appears both in the ambient space and the moduli space of SW curve.
\item[(e)] After passing Argyres-Douglas loci of (d), we again have 5 separated singular points on the $u$-plane. Especially $u_{0}^{Q}$ and $u_{2}^{C}$ are separated. 
However note that the BPS dyon charges of vanishing cycles changed  from the cases of (a), (b), and (c) given in Eqn. \eqref{cycle} as we go through cusp-like (or Argyres-Douglas) singularity of (d). 
Instead of $\nu^Q_2$, we have a new vanishing cycle
\begin{equation}
 \nu_{2^\prime}^{Q}=-\beta_{2}-\alpha_{1}-2\alpha_{2} = \nu_{0}^{Q}+\nu_{2}^{Q}. \label{chargeJump}
 \end{equation}

\end{description}
    
\begin{figure}[h]
        \begin{center}
\includegraphics[width=.5\textwidth]{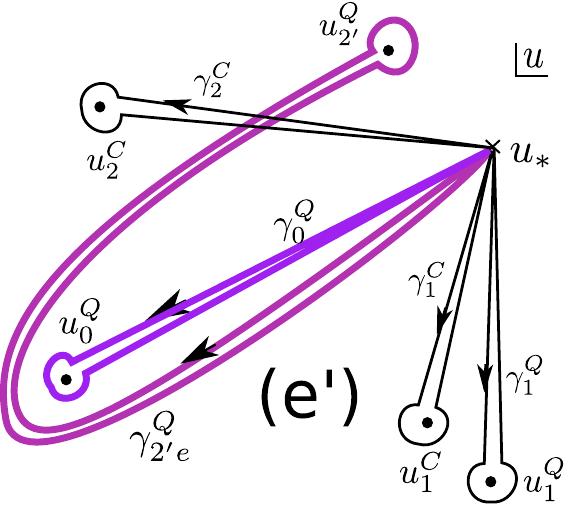}
        \end{center}
        \caption{Alternative choice for trajectory in moduli slice}\label{eprime}
        \end{figure}

In observing the right side of Fig. \ref{c2_10pic_down}, we realize that the set of vanishing 1-cycles have changed in going from (c) to (e), through (d). 
     In order to check whether this ``jump'' is something intrinsic - geometrically and physically - to this system, we can conduct a few tests. One quick test we perform here is to choose alternative paths on moduli slice and check how it may affect vanishing 1-cycles. Instead of the paths on the left of (e) in Fig. \ref{c2_10pic_down}, we can change a new set as given in Fig. \ref{eprime}.

         The new path still surrounds the same singularity, however its relative location with respect to other singularities changed. Two trajectories in the moduli space $\gamma^Q_{2^\prime}$ in Fig. \ref{c2_10pic_down} and $\gamma^Q_{2^\prime e}$ in Fig. \ref{eprime} both surround only $u^Q_{2^\prime}$, a singular point, on the moduli slice. In other words, they form non-contractible loops surrounding a singular locus $\Sigma^Q_{2^\prime}$ in Fig. \ref{moduli3dcolor} and nothing else. However the key difference is their relative location with respect to two other singularities $\Sigma^Q_{0}$ and $ \Sigma^C_{2}$ in Fig. \ref{moduli3dcolor} (or $\gamma^Q_{0}$ and $ \gamma^C_{2}$ in Fig. \ref{c2_10pic_down} and Fig. \ref{eprime}. 
    
\begin{figure}[h]
        \begin{center}
\includegraphics[width= \textwidth]{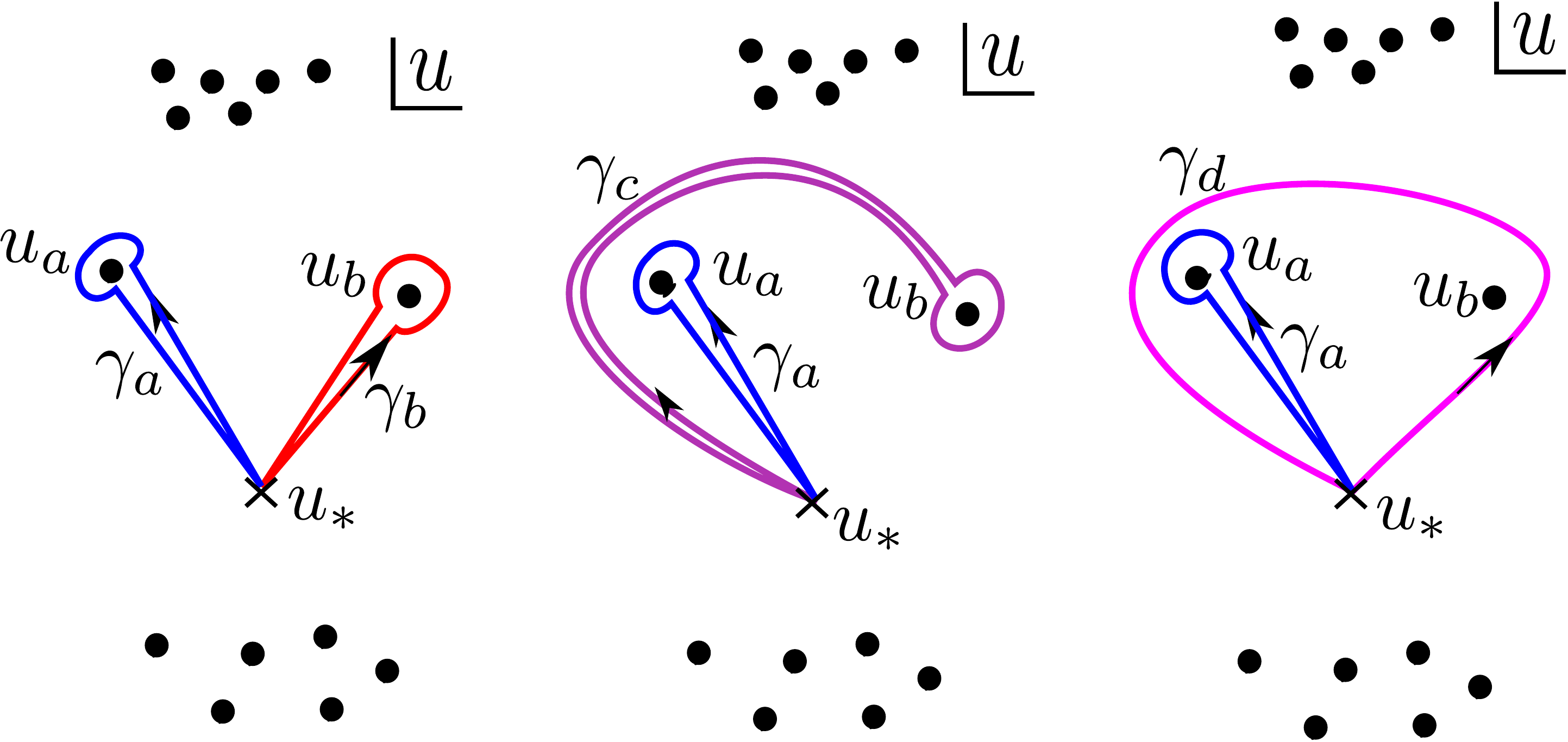}
        \end{center}
        \caption{Different trajectories in moduli space surrounding same singular point}\label{abc}
        \end{figure}

In order to answer how this change may affect the assignment of dyon charges of vanishing 1-cycles, let us consider following cartoon in Fig. \ref{abc}. 
On the same moduli slice, we have a few different trajectories drawn, surrounding singular points of vanishing discriminant of the curve. We have drawn two other bunches of singular points in the top and the bottom, to denote other singularities whose spatial relationship with all the trajectories depicted remain unchanged. Both $\gamma_b$ and $\gamma_c$ surround a singular point $u_b$ only. We can associate a monodromy matrix for each closed loop in the moduli space, to denote how the 1-cycles get transformed among themselves, in the spirit of Picard-Lefshetz formula given in Eqn. \eqref{PLformula}. 
Trajectories $\gamma_b$ and $\gamma_a$ can be combined, in that order, to give a trajectory $\gamma_d$, and similarly trajectories $\gamma_a$ and $\gamma_c$ can be combined to give a trajectory $\gamma_d$. 
Therefore we have the following relations among the monodromy matrices (The matrix on the right will be operated first.) 
\begin{equation} M_a M_b = M_c M_a = M_d.  \label{Mmulti} \end{equation}

%
%
%
%
%

For a SW curve of genus 1, the monodromy matrix is given in Ref. \refcite{LercheReview} as
\begin{equation}
M^{(g,q)}=\left( \begin{tabular}{ cc }  $1+qg$  & $q^2$ \\$-g^2$ &  $ 1-gq $ \end{tabular} \right) \label{monomat}
\end{equation}
for a closed loop in moduli space, which surrounds a singularity associated with vanishing 1-cycle $g \beta + q \alpha$ (or a massless dyon of charge $(g,q)$). If dyon has vanishing charge or if dyon is absent, then the monodromy matrix reduces to the identity.

For higher genus case which we are considering here, we can arrange the dyon charges to contain only $\alpha_1, \beta_1$ with vanishing contribution from other $\alpha_i$'s and $\beta_i$'s, by symplectic transformation (or electromagnetic dualities). In other words, the dyon charge will look like 
\begin{eqnarray}
(\vec{g},\vec{q})& = &(g_1,g_2, \ldots, g_r; q_1,q_2, \ldots, q_r) \nonumber \\ 
&=& (g_1,0, \ldots, 0; q_1,0, \ldots, 0) = (g,0, \ldots, 0; q,0, \ldots, 0) = g\beta_1 + q \alpha_1
\end{eqnarray}
This is equivalent to saying that 
we can arrange the monodromy matrices $M_a$ and $M_b$ of Fig. \ref{abc} to take the following form 
\begin{equation}
M^{(g,0, \ldots, 0; q,0, \ldots, 0)}=\left( \begin{tabular}{ cc |c  }  $1+qg$  & $q^2$ & $0$ \\$-g^2$ &  $ 1-gq $ & $0$ \\ \hline $0$ & $0$& $0$ \end{tabular} \right), \label{monomatg}
\end{equation}
which contains Eqn. \eqref{monomat} as a left-top $2 \times 2$ block. All the $0$'s in Eqn. \eqref{monomatg} are to be understood as submatrices of appropriate sizes. 
  In the remaining part, we will omit the $0$'s and write down only the first $2 \times 2$ submatrix as the monodromy matrices.

Again by symplectic transformation with no loss of generality, we can assign the dyon charges of $\gamma_a$ and $\gamma_b$ to be $(0,a)$ and $(b,0)$, with intersection number 
\begin{equation}
(0,a) \circ (b,0) = a \alpha \circ b \beta = -a b .
\end{equation} Please note we assign the dyon charges to the trajectories $\gamma$'s but not to the singular points $u$'s themselves.

From Eqn. \eqref{Mmulti} and Eqn. \eqref{monomat}, we have monodromy matrices as given below:
 \begin{eqnarray} 
   M_a &=&M^{ (0,a)}= \left( \begin{tabular}{ cc }  $1$  & $a^2$ \\$0$ &  $ 1$ \end{tabular} \right)=  \left( \begin{tabular}{ cc }  $1$  & $-a^2$ \\$0$ &  $ 1$ \end{tabular} \right)^{-1}, \nonumber \\
 M_b &=& M^{ (b,0)} = \left( \begin{tabular}{ cc }  $1$  & $0$ \\$-b^2$ &  $ 1$ \end{tabular} \right) =\left( \begin{tabular}{ cc }  $1$  & $0$ \\$b^2$ &  $ 1$ \end{tabular} \right)^{-1}, \nonumber \\
  M_c & = & M_a M_b M_a^{-1} =\left( \begin{tabular}{ cc }  $1-a^2 b^2$  & $a^4 b^2$ \\$-b^2 $ &  $ 1+a^2 b^2$ \end{tabular} \right) = M^{ \pm (b,- a^2 b)}.   \label{Mcshift}
\end{eqnarray}  
Note that there is an ambiguity for the overall sign of dyon charge associated to $M_c$, because Eqn. \eqref{monomat} is invariant under $(g,q) \rightarrow (-g,-q)$.

For fun, we can also examine $M_d$ and attempt (and fail) to interpret it in terms of dyon charges. 
From Eqn. \eqref{Mmulti}, monodromy matrix for $\gamma_d$ is given as
\begin{equation} M_d  = M_a M_b  = \left( \begin{tabular}{ cc }  $1-a^2 b^2$  & $a^2$ \\$-b^2$ &  $ 1$ \end{tabular} \right)
\label{Md} \end{equation} whose trace matches that of Eqn. \eqref{monomat} only for $ab=0$. Therefore $M_d$ cannot be interpreted as a singularity of a single vanishing 1-cycle for $ab \ne 0$.

We can interpret the result of Eqn. \eqref{Mcshift} and Fig. \ref{abc} as following: 
\begin{enumerate}
\item The dyon charge of a singular point $u_b$ depends on choice of trajectory in the moduli slice. 
\item If the trajectory passes through another singular point $u_a$ (but not surrounding it by a closed loop), then dyon charge of $u_b$ gets shifted by multiple of dyon charge of $u_a$. 
\item If we chose the first sign in Eqn. \eqref{Mcshift}, then this relation becomes \begin{equation}
(b,0) \rightarrow (b,0) - \left[ (b,0) \circ (0,a) \right] (0,a) = (b, - a^2 b),  \label{abrel}
\end{equation} with a close agreement with 
  Eqn. \eqref{PLformula}.
\end{enumerate}
We can now write down general formula for how dyon charge gets changed. In Fig. \ref{abc}, for each trajectories $\gamma_{a}, \gamma_{b}, \gamma_{c}$, let us say vanishing 1-cycles are $\nu_{a}, \nu_b, \nu_{c}$. Then their relationship is given as similar to Eqn. \eqref{abrel} as
\begin{equation}
\nu_c = \nu_b - ( \nu_b \circ \nu_a )  \nu_a. \label{abcmono}
\end{equation}

Going back to the question of spectra jump for $Sp(4)$ related to Fig. \ref{eprime}, now we can use the techniques we learned above, especially that of Eqn. \eqref{abcmono}.
First, by inspecting Fig. \ref{eprime} and Fig. \ref{abc}, we can plug in $\nu_a = \nu_0^Q$, $\nu_b= \nu_{2^\prime}^Q$, $\nu_c = \nu_{2^\prime e}^Q$ into Eqn. \eqref{abcmono} to obtain 
\begin{equation}
\nu_{2^\prime e}^Q =\nu_{2^\prime}^Q - ( \nu_{2^\prime}^Q \circ \nu_0^Q )  \nu_0^Q = \nu_{2^\prime}^Q  -  \nu_0^Q = \nu_{2}^{Q}, \label{22mono}
\end{equation}
where the last equality was given from Eqn. \eqref{chargeJump}. 
If somehow we can argue that $\gamma^Q_{2^\prime e }$ of Fig. \ref{eprime} is more natural choice of trajectory in moduli space than $\gamma^Q_{2^\prime   }$ of Fig. \ref{c2_10pic_down}, then we can undo the spectra jump as we move from (c) to (e) in Fig. \ref{moduli3dcolor}. 
It may suggest that we can avoid the spectra jump (in dyon charges of massless states) if we choose a different path on moduli slice.
However for the moment we cannot say conclusively what will be the most natural and consistent choice of trajectory paths. It will be interesting to check this by focussing on massless sector of wall-crossing formulas of BPS spectra given in Refs. \refcite{GMNwall,cordovafa,cordovafa0}. 
  
\subsection{Maximal Argyres-Douglas theories and dual Coxeter number \label{maxADh}}

So far we considered Argyres-Douglas theories with two massless dyons. Here, we will consider {\it maximal} Argyres-Douglas theories, where maximal number of dyons become massless and are mutually non-local. This is achieved by bringing the maximal number of branch points on $x$-plane, so that $f(x)$ will have a root with maximal degeneracy. Let us recall the SW curve for $SU(r+1)$ gauge group given Eqn. \eqref{surcurve1form} and Eqn. \eqref{fpm}. 
\begin{equation}  
 y^{2}=f_{SU(r+1)} = f_{+} f_{ -} , \quad   
  f_{\pm} \equiv  x^{r+1} + \sum_{i=1}^r u_i x^{r-i}   \pm \Lambda
^{ r+1}.  
\end{equation}   
If we have $u_r = \mp \Lambda^{r+1}$ while all other $u_i$'s vanish, then $f_{\pm} =x^{r+1}$ holds and $f(x)$ has a root $x=0$ with maximal degeneracy\cite{ArgyresDouglas, LercheReview}. It is straightforward to find 2 maximal Argyres-Douglas points in moduli space of pure $SU(r+1)$ and $SO(2r)$ SW theories. 

Taking no-flavor limit of Ref. \refcite{ArgyresShapere}, we have SW curve for pure $SO(2r)$ as 
\begin{equation} y^2 = C_{SO(2r)}^2 - \Lambda^{2(2r-2 )} x^4 =C_{SO(2r),+} C_{SO(2r),-}   \end{equation}
 with
 \begin{eqnarray} C_{SO(2r),\pm} & =& C_{SO(2r) }  \pm \Lambda^{ (2r-2 )} x^2, \nonumber \\ 
  C_{SO(2r)}(x) & \equiv  &  x^{2r} + s_2 x^{2r-2} +\cdots + s_{2r-2} x^2 +\tilde{s}_r^2,
  \end{eqnarray}
  in agreement with Ref. \refcite{KLYTsimpleADE}.
Some of the monodromy properties for pure $SO(2r)$ were studied in Ref. \refcite{BLSo} with emphasis on the $SO(8)$ example. 

Maximal Argyres-Douglas points for the $SO(2r)$ will be two moduli points given by \cite{EHIY}
\begin{equation}
 s_{2r-2} =  \Lambda^{ \pm (2r-2 )}, \quad s_{2i}=0, ~~i\ne r-1 \label{SOevenMaxAD}
 \end{equation}
 which makes $C_{SO(2r),\mp} = x^{2r}$ to have a root with maximal order of vanishing. Just as in the $SU(r+1)$ case, this computation is straightforward, partially thanks to $\mathbb{Z}_2$ symmetric structure between $C_{SO(2r),\pm}$ and between $f_\pm$ in the SW curve, which is lacking in the $Sp(2r)$ and $SO(2r+1)$ cases.
Scaling behavior at maximal Argyres-Douglas points for $SU(r+1) $ and $SO(2r)$ SW theory was studied in Ref. \refcite{AMT} recently, with focus on the shape of moduli space in the neighborhood of those theories. More specifically, the quantum Higgs branch appears\footnote{Some of the best place to learn about Higgs branch are Refs. \refcite{AlvarezGaumeHassanReview,Argyres:1998vt}.}.

Locating maximal Argyres-Douglas (AD) points for $Sp(2r)$ and $SO(2r+1)$ SYM involves more algebra, and the number of maximal Argyres-Douglas points equals to the dual Coxeter number of the gauge group.
For pure $Sp(2r)$ theory, Ref. \refcite{DSW} proposes $r+1$ candidates for maximal Argyres-Douglas theories, while Ref. \refcite{SD} proposes $2r-1$ candidates for maximal Argyres-Douglas points of pure $SO(2r+1)$ theory.

Recall, from Eqn. \eqref{sprcurve1form} and Eqn. \eqref{fCfQdef}, the SW curve for $Sp(2r)$ SYM is
\begin{equation}    
 y^{2}=f_{Sp(2r)}= f_C f_Q , \quad 
f_C =  x^r + \sum_{i=1}^r u_i x^{r-i}   , \quad f_Q \equiv  x f_C  +  16 \Lambda^{2r+2}.  
\end{equation}   
Maximal AD points occur when we bring all roots of $f_Q=0$ together. This happens at $r+1$ points in moduli space, where the curve develops $A_{r}$ singularity\cite{DSW}.
   This occurs when the moduli take the following values: 
   \begin{align} 
u_i=&   \left(
\begin{array}
[c]{c}%
r+1\\
i
\end{array}
\right)  (-Q)^{i},  
\end{align}
with $Q$ given by 
\begin{equation}
Q=- \exp\left(  \frac{2\pi i}{r+1}k\right)  \left(  16\right)  ^\frac{1}{r+1} \Lambda^{2}, \qquad k\in \mathbb{Z}. \label{Qvalue}
\end{equation}
This forces the $f_Q$ to have a root with maximal degeneracy as
\begin{equation}
f_Q=(x-Q)^{r+1} . \label{fQQ}
\end{equation} 
It also follows that $C_i$'s, the roots of $f_C$ are given as below: 
\begin{equation}
\{ C_i \}=  \left\{  Q\left(  1-\exp\left(  \frac{2\pi i}{r+1}k\right)  \right)  \right\} , \qquad k \in \mathbb{Z}, \qquad k \notin (r+1) \mathbb{Z}. \label{Cvalues}
\end{equation}

  Given that Eqn. \eqref{Qvalue} allowed for $r+1$ different choices of phase for $Q$, we have $\mathbb{Z}_{r+1}$ symmetric $r+1$ points on the moduli space, where maximal Argyres-Douglas singularity occurs.

Similarly, $2r-1$ maximal Argyres-Douglas points of 
 $SO(2r+1)$ SYM were located in Ref. \refcite{SD}. 
 The curves for 
$SO(2r+1)$ and $SO(2r)$ SYM are almost similar, but $SO(2r+1)$ case is much harder to solve for the maximal Argyres-Douglas points. 

We again take no-flavor limit of the curve of Ref. \refcite{ArgyresShapere} to obtain the curve for pure SW theory\cite{KLYTsimpleADE,SO5DS},  
 \begin{equation}
  y^2 = f_{SO(2r+1) }(x)=C_{SO(2r+1) }^2 - \Lambda^{2(2r-1 )} x^2  = C_{SO(2r+1) ,+}  C_{SO(2r+1), -}  \label{SOoddcurve}  
  \end{equation}
with 
\begin{eqnarray} C_{SO(2r+1) ,\pm} &=& C_{SO(2r+1) }  \pm \Lambda^{ (2r-1 )} x ,\nonumber \\
C_{SO(2r+1) }(x) &\equiv&  x^{2r} + s_2 x^{2r-2} +\cdots + s_{2r-2} x^2 + {s}_{2r}.
\end{eqnarray}

Two polynomials $C_{SO(2r+1) ,\pm}$ will share a root if $x=0$. This, however, does not give much mileage for bringing maximal number of branch points together for $f_{SO(2r+1) }(x)$. This root does not have high enough order of vanishing. Instead we will work on having $C_{SO(2r+1), +}$ (or equivalently $C_{SO(2r+1) , -}$) to have a root with maximal order of vanishing. Since we have only $r$ degrees of freedom, we can only bring together $r+1$ branch points, and the maximal order of vanishing is $r+1$. The best we can do is to bring $C_{SO(2r+1), +}$ into the following form:\cite{SD}
 \begin{equation}
 C_{SO(2r+1), +} = (x+b)^{r+1} (x^{r-1} + u_1 x^{r-2} + u_2 x^{r-3} + \cdots + u_{r-2} x + u_{r-1} ) \label{CmaxCusp}
 \end{equation}
when the moduli $s_{2i}$'s satisfy 
\begin{equation}
s_{2k} = (-b^2)^k \frac{(2r-1)}{(2r-2k-1)} \frac{r!}{k! (r-k)!}, \label{svalueformAD}
\end{equation}    
  with $b$ given by \begin{equation}   b   = \omega_{2r-1}^k \left[ (-1)^{r+1} \frac{(2r-3)!!}{(2r)!!} \right]^{1/(2r-1)}\Lambda , \quad k\in \mathbb{Z}. \label{bvalueformAD}
 \end{equation}
   Here $\omega_m$ is $m$'th root of unity. Eqn. \eqref{bvalueformAD} allows $2r-1$ possible values of $b$. In turn there are $2r-1$ solutions to Eqn. \eqref{svalueformAD}. Therefore we have $2r-1$ isolated points, which are $\mathbb{Z}_{2r-1}$ symmetric among themselves, in the moduli space of pure $SO(2r+1)$ SYM where maximal Argyres-Douglas theory occurs.

\section{D\'{e}j\`{a} Vu: Singularity Tools: Exterior Derivative \& Double Discriminant  \label{doublediscsection}}  
So far we encountered a few tools to detect singularity of the SW curves. In subsection \ref{GeomReview}, we discussed various tools for singularity search, namely exterior derivative $d$ in ambient space, and discriminant. In Eqn. \eqref{doublediscsp4}, we saw that considering double discriminant $\Delta_u \Delta_x f(x)$ gives candidates for interesting singularities with 2 massless dyons. 
 Now let us write down general rules in a systematic way. 

We detected singularity of a Riemann surface embedded in the ambient space whose coordinates are $x, y$ by taking an exterior derivative, or by demanding all the partial derivatives with respect to $x$ and $y$ to vanish. 
Similarly, when an algebraic variety in moduli space forms a singularity, it is captured by demanding the exterior derivative to vanish inside the moduli space. 
Recall that the vanishing discriminant loci of the SW curve form a complex codimension 1 algebraic variety inside the moduli space, which is 
given by $\Delta_x f =0$. Therefore it follows that $d \Delta_x f =0$ (where $d$ is taken inside the moduli space) will pinpoint us to where $\Delta_x f =0$ forms a singularity, namely where discriminant loci intersect themselves. That is where we have multiple vanishing 1-cycles (massless BPS dyons). 

 As explained in subsection \ref{GeomReview}, the exterior derivative $d$ can be written in terms of the partial derivatives with respect to all the coordinates of the ambient space. Therefore the exterior derivative inside the moduli space is given as $d= \sum_{i=1}^{r} du_i \frac{\partial  }{\partial u_i} $.
 One might think that demanding the $d =0$ actually reduces $r$ degrees of freedom, since we demand all the $r$ partial derivatives to vanish. However, $ \Delta_x f=d \Delta_x f=0$ indeed contains codimension 2 solutions (instead of codimension $r+1$).

When the $ \Delta_x f=0$ loci become singular (where  $ \Delta_x f=d \Delta_x f=0$ holds), double discriminant also vanishes ($\Delta_u \Delta_x f(x)=0$). In other words, singularity is seen from the viewpoint of moduli slices (parallel $u$-planes) as well. However the converse does not hold. Something might appear singular on certain moduli slices but it can be smooth in the full ambient space (moduli space).  
 In other words, $ \Delta_x f= \Delta_u \Delta_x f=0$ is a necessary but not sufficient condition for having $ \Delta_x f=d \Delta_x f=0$. 
  
    \subsection{Exterior derivative detects coexistence of multiple massless BPS dyons \label{exteriordVSdd}}
 The
vanishing discriminant condition of the SW curve $\Delta_x f=0$ defines an algebraic variety $\Sigma$. Since $\Delta_x f$ is written only in terms of moduli $u_i$'s (without $x$ and $y$), $\Sigma$ is an algebraic variety embedded inside the moduli space, denoting moduli loci of massless BPS states. When this algebraic variety $\Sigma$ self-intersects, two or more BPS states become massless, which occurs when we demand $\Delta_x f=d \Delta_x f =0$.  

\begin{figure}[h]
\begin{center}
\includegraphics[
width=2.5in]{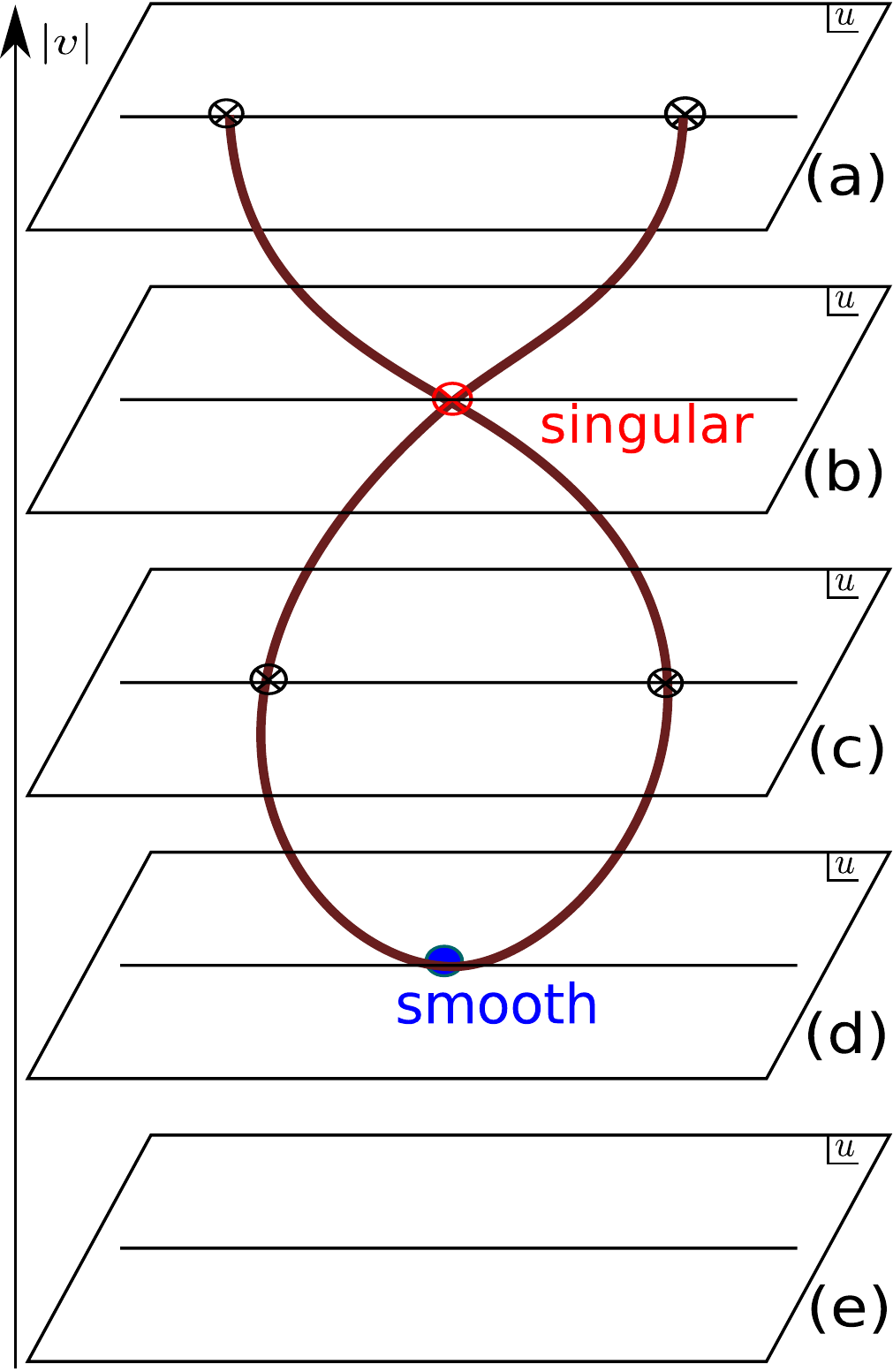}
\end{center}
\caption{A heuristic example showing difference between vanishing double discriminant and a vanishing exterior derivative. Vanishing double discriminant will single out slices (b) and (d), while the exterior derivative will dictates that only (b) is a singularity. Vanishing double discriminant is a necessary but not a sufficient condition for a singularity. Compare this picture with {Fig.~\ref{moduli3dcolor}}.}
\label{figure8}
\end{figure}
 
An heuristic example is depicted 
in {Fig.~\ref{figure8}}. Inside a moduli subspace, we draw a figure-eight-like object, which is analogous to $\Delta_x f=0$ loci as in {Fig.~\ref{moduli3dcolor}}. We mark various $u$-planes with (a) to (e). The number of singular points changed on each $u$-planes. When the singular points collide on the $u$-plane we have $\Delta_u\Delta_x f=0$, for example on slices (b) and (d). Slice (b) is true singularity, while (d) is not. Thus we see 
that $\Delta_x f=\Delta_u \Delta_x f =0$ is a necessary but not sufficient condition to have $\Delta_x f=d \Delta_x f =0$.

Fig.~\ref{moduli3dcolor} and Fig.~\ref{figure8} 
look very similar to each other in that it displays the singularity structure inside moduli space. Both shows the collision of discriminant loci and formation of higher singularity - 
(b) and (d) of Fig.~\ref{moduli3dcolor} and (b) of Fig.~\ref{figure8}. The main difference is this: Fig.~\ref{figure8} contains an example where $\Delta_x f=\Delta_u \Delta_x f =0$ holds but $d \Delta_x f \ne 0$ on its part (d). 

\subsection{Factorization of double discriminant, and order of vanishing  \label{FactorDD}} 
 
Massless dyons coexist at complex-codimension-2 loci where both discriminant and its exterior derivative vanish. There the curve looks like either of following two:\cite{SD}
\begin{itemize}
\item[{\bf Cusp}] 
The curve $y^2=(x-a)^3 \times \cdots  $ has a cusp-like singularity. Vanishing discriminant $\Delta_x f=0$ locus also intersects with  {{}cusp}-like singularity in moduli space. 
There $\Delta_u \Delta_x f=0$ also holds, with order of vanishing 3. 
Two massless dyons are mutually non-local. \item[{\bf Node}] The curve $y^2=(x-a)^2 (x-b)^2 \times \cdots  $ has a node-like singularity.  
Vanishing discriminant $\Delta_x f=0$ locus also intersects with {{}node}-like singularity. There $\Delta_u \Delta_x f=0$ also holds, with order of vanishing 2.  Two massless dyons are mutually local.
\end{itemize}
  
%

 Order of vanishing of each root of $\Delta_u \Delta_x f$ tells us the type of singularities. 
In order to justify that, 
we discuss roots of $\Delta_u \Delta_x f=0$. 
When $ \Delta_x f=0$ and $d \Delta_x f=0$ hold, each root of vanishing double discriminant $\Delta_u \Delta_x f=0$ corresponds to two massless dyons with appropriate combinatoric meaning.  
  
Double discriminants of $SU(r+1)$ and $Sp(2r)$ factorize as:\cite{SD}
  \begin{eqnarray}   
  \Delta_u \Delta_x f_{SU(r+1)}= \Delta_u \Delta_x (f_+ f_-) &=&\# \left( v^{(2r+2)^2} + \cdots \right) \nonumber\\  
 &=& \#  \left( v^{(r+1)^2} + \cdots \right)^2  \left( v^{r+1} + \cdots \right)^3 \left( v^{r+1} + \cdots \right)^3 \nonumber\\  & &\times  
 \left( v^{(r+1)(r-2)/2} + \cdots \right)^2 \left( v^{(r+1)(r-2)/2} + \cdots \right)^2 \nonumber\\
 & \equiv & \# ({PN_{II}})^2 ({N_{III}})^3 ({P_{III}})^3 ({N_{II}})^2 ({P_{II}})^2,
 \label{DDsu}
  \end{eqnarray}   
  \begin{eqnarray}   
 \Delta_u \Delta_x f_{Sp(2r)} =  \Delta_u \Delta_x(f_Q f_C)   &=&\# \left( v^{(2r+1)^2} + \cdots \right) \nonumber\\   
 &=& \#  \left( v^{r(r+1)} + \cdots \right)^2   \left( v^{r} + \cdots \right)^3 \left( v^{r+1} + \cdots \right)^3 \nonumber\\  & &\times  
 \left( v^{r(r-3)/2} + \cdots \right)^2 \left( v^{(r+1)(r-2)/2} + \cdots \right)^2 \nonumber\\
 & \equiv & \# ({QC_{II}})^2 ({C_{III}})^3 ({Q_{III}})^3 ({C_{II}})^2 ({Q_{II}})^2, \label{DDsp} \end{eqnarray}    
where $u\equiv u_1, v\equiv u_2$ denote the two moduli among $r$ complex moduli. 
  The subscripts $II$ and $III$ denote the order of vanishing - each corresponding to node and cusp like singularity.

\vskip.1in

\noindent {\bf Case I:} ~ As an example, the first factor in second line of Eqn. \eqref{DDsu} is 
     \begin{equation} ({PN_{II}}) \equiv \left( v^{(r+1)^2} + \cdots \right). \label{pn2} \end{equation}
     This corresponds to having two pairs of branch points on the $x$-plane collide each other pairwise, where each pair is $P_i$ type and $N_i$ type respectively. The curve degenerates into a node-like singularity
     \begin{equation} y^2 = (x-P_i)^2 (x-N_j)^2  \times \cdots .\end{equation}
Number of choices for choosing one pair of $P_i$'s and $N_i$'s is given as: 
\begin{equation}
\left(
\begin{array}
[c]{c}%
r+1\\
1
\end{array}
\right)^2 = { (r+1)^2},
 \end{equation} which is exactly the power of $v$ in \eqref{pn2}.  
 
\vskip.1in 

\noindent {\bf Case II:}~
The first factor in the third line of Eqn. \eqref{DDsu} is
  \begin{equation}  ({N_{II}}) \equiv \left( v^{{{}(r+1)(r-2)/2}} + \cdots \right), \end{equation} 
and this corresponds to the scenario where two pairs of $N$-type branch points on the $x$-plane collide each other pairwise. The curve degenerates into a node-like singularity
    \begin{equation}  y^2 = (x-{N_i})^2 (x-{N_j})^2  \times \cdots  .\end{equation} 
     
\vskip.1in

\noindent {\bf Case III:}~
The second factor in the second line of Eqn. \eqref{DDsu}  
  \begin{equation}  ({N_{III}}) \equiv \left( v^{r+1} + \cdots \right) \end{equation} 
is related to having three $N$-type branch points on the $x$-plane collide all together. The curve degenerates into a cusp-like singularity
     \begin{equation}  y^2 = (x-{C_i})^{3}  \times \cdots  . \end{equation} 
     
     \vskip.1in

  Similarly, we can understand other factors of Eqn. \eqref{DDsu} and Eqn. \eqref{DDsp}.

\section{Conclusion \label{conclusion}}


In this review, we observed the relevance and importance of supersymmetric Yang-Mills theories. We discussed ${\cal N}=2$ SYM with classical gauge groups, with particular attention to their singularity structure associated with massless states. We translated the physics questions into the language of Seiberg-Witten geometry, so that a hyperelliptic curve equipped with a 1-form encodes physical information of SYM.

 \begin{equation}
\begin{tabular}{|ccc|}
\hline Physics & $\leftrightarrow$ &  Geometry  \\ \hline
\hline  supersymmetric Yang-Mills theory & $\leftrightarrow$ & Hyperelliptic curve  \\  \hline
rank of gauge group  & $\leftrightarrow$ & genus  \\ \hline
particle  & $\leftrightarrow$ &  1-cycle \\ \hline
 mass &$\leftrightarrow$ & $ \left| \oint \lambda \right| $    \\ \hline
 massless states  & $\leftrightarrow$ & singularity  \\
  & & (vanishing 1-cycle at $\Delta_x f=0$) \\  \hline
 massless & $\leftrightarrow$ &   worse singularity \\
  $e^-$ \& magnetic monopole  & &  such as cusp  \\
\hline  \end{tabular} 
\end{equation}

With this dictionary in mind,
we examined singularity loci of families of hyperelliptic curves which are associated with pure SW theories. 
At discriminant loci $\Delta_x f=0$ of the SW curve, we have vanishing 1-cycles. We identified BPS dyon charges of all the 
$2r+1$ and $2(r+1)$ vanishing cycles respectively for pure $Sp(2r)$ and $SU(r+1)$ SW curves.   

When discriminant loci form a singularity inside the moduli space ($d \Delta_x f=0$), multiple massless dyons coexist. 
Here the `double discriminant' also vanishes ($\Delta_u \Delta_x f=0$). Note however that the converse does not hold.
If order of vanishing of roots to the double discriminant is high (equal to $3$), then vanishing 1-cycles coexist and intersect: we are at the Argyres-Douglas loci. 
  
On top of many open questions proposed in Ref. \refcite{SD}, it will be interesting to study the behavior of the SW curve near the Argyres-Douglas loci, extending the works of Ref. \refcite{AMT}, which discusses appearance of quantum Higgs branch at maximal Argyres-Douglas points of SW theories. It will be also interesting to understand the results reviewed here in the context of wall-crossing and quiver mutation in Refs. \refcite{GMNwall,cordovafa,cordovafa0}. Finally, the analysis of monodromy may benefit from making more connection to the braiding procedure in knot theory.

\section*{Acknowledgments}
 
It is a great pleasure to thank Murad Alim, Philip Argyres, Heng-Yu Chen, Keshav Dasgupta, Hoyun Jung, Dong Uk Lee, Andy Neitzke, Jihun Park, Alfred Shapere, Yuji Tachikawa, Donggeon Yhee, and Philsang Yoo for helpful discussions. Long Chen, Pedro Liendo, Chan-Youn Park, and especially Philsang Yoo gave invaluable feedback on the manuscript. The author benefited from encouragement from Howard, Mini, Sun-young Park, and Daniel Tsai. This review is dedicated to the memory of Arthur G.

 The work is supported in
part by NSERC grants. All the figures are created by the author using the program Inkscape with TeXtext.

\bibliographystyle{ws-ijmpa} 
\bibliography{AScurve}
\end{document}